\documentclass[preprint,notitlepage,nofootinbib]{revtex4}
\usepackage{ulem}
\usepackage{amssymb, amsmath, amsopn, color, graphicx}
\usepackage[english]{babel}

\numberwithin{equation}{section}
\renewcommand\theequation{\arabic{section}.\arabic{equation}}

\usepackage{bm}
\DeclareMathAlphabet{\matheul}{U}{eus}{m}{n}
\usepackage{float}
\newcommand{\del}{\nabla}
\usepackage[colorlinks]{hyperref}
\hypersetup{linkcolor=blue,citecolor=blue,filecolor=black,urlcolor=blue}

\newcommand{\diag}{\mathop{\mathrm {diag}}}
\newcommand{\sign}{\mathop{\mathrm {sign}}}
\newcommand{\tr}{\mathop{\mathrm {tr}}}
\newcommand{\zzbar}{$Z\bar{Z}$}

\newcommand{\tK}{{\widetilde{K}}} % the azimuthal Killing vector
\newcommand{\sd}{\varepsilon_{\mkern-2mu{}_{\Delta}}} % \sign(\Delta)
\newcommand{\muo}{\hat{\mu}_{0}} % rotating T^t_{~t}
\newcommand{\mui}{\hat{\mu}_{1}} % rotating T^r_{~r}
\newcommand{\muii}{\hat{\mu}_{2}} % rotating T^\theta_{~\theta}
\newcommand{\muiii}{\hat{\mu}_{3}} % rotating T^\phi_{~\phi}
\newcommand{\muiiio}{\hat{\mu}_{30}} % rotating T^t_{~\phi}
\newcommand{\muIii}{\hat{\mu}_{12}} % rotating T^r_{~\theta}
\newcommand{\DIii}{\hat{D}_{12}} % discriminant rtheta block
\newcommand{\Diiio}{\hat{D}_{30}} % discriminant tphi block
\newcommand{\BIii}{\hat{B}_{12}} % trace rtheta block
\newcommand{\Biiio}{\hat{B}_{30}} % trace tphi block
\newcommand{\siiio}{\hat{\sigma}_{30}} % rotating T^t_{~\phi}
\newcommand{\sIii}{\hat{\sigma}_{12}} % rotating T^t_{~\phi}

\newcommand{\Bigbra}{{\Big[\mkern-6mu\Big[}}
\newcommand{\Bigket}{{\Big]\mkern-6mu\Big]}}

\allowdisplaybreaks

\begin{document}

%\preprint{APS/123-QED}

\title{Physical interpretation of Newman-Janis rotating systems.\\ II. General systems}% Force line breaks with \\
\thanks{}%

\author{Philip Beltracchi}
\email{phipbel@aol.com}
\affiliation{Department of Physics and Astronomy, University of Utah\\Salt Lake City, Utah 84112}

\author{Paolo Gondolo}
\email{paolo.gondolo@utah.edu}
\affiliation{Department of Physics and Astronomy, University of Utah\\Salt Lake City, Utah 84112}
\affiliation{Department of Physics, Tokyo Institute of Technology, 2-12-1 Ookayama, Meguro-ku, Tokyo 152-8551, Japan}
\affiliation{Kavli Institute for the Physics and Mathematics of the Universe, The University of Tokyo, Kashiwa, Chiba 277-8583, Japan}

\collaboration{}%\noaffiliation


\begin{abstract}
\noindent
Drake and Szekeres have extended the Newman-Janis algorithm to produce stationary axisymmetric spacetimes from general static spherically symmetric solutions of the Einstein equations. The algorithm mathematically generates an energy-momentum tensor for the rotating solution, but the rotating and nonrotating system may or may not represent the same physical system, in the sense of both being a perfect fluid, or an electromagnetic field, or a $\Lambda$-term, and so on. In Part I (arxiv:2104.02255), we compared the structure of the eigenvalues and eigenvectors of the rotating and nonrotating energy-momentum tensors (their Segre types) and looked for the existence of equations of state relating the rotating energy density and principal pressures for Kerr-Schild systems. Here we extend our analysis to general static spherically symmetric systems obtained according to the Drake-Szekeres generalization of the Newman-Janis algorithm. We find that these rotating systems can have almost all Segre types except [31] and [(31)]. Moreover, the Segre type of the spacetime can change severely in passing from the nonrotating to the rotating configurations, for example to $[11Z\bar{Z}]$ from seed systems which were initially [(111,1)]. We also find conditions dictating how many equations of state may exist in a Drake-Szekeres system.
\begin{description}
\item[Usage]
\item[PACS numbers]
\item[Structure]
\end{description}
\end{abstract}

\pacs{Valid PACS appear here}% PACS, the Physics and Astronomy
\maketitle

%\tableofcontents
\section{Introduction}

Within general relativity, one may find rotating solutions starting from nonrotating spherically symmetric solutions utilizing the Newman Janis algorithm and its generalizations.  The original algorithm allowed for a rederivation of the Kerr solution and initiated the discovery of the Kerr-Newman solution \cite{Newman:1965tw,Newman:1965my,Kerr:2007dk}, by starting with the Schwarzschild and Riessner-Nordstrom spherical solutions. The algorithm was extended by Gurses and Gursey to generate rotating systems from spherically symmetric solutions of the Kerr-Schild type \cite{Gurses:1975vu}, such systems have seen extensive  use in recent years modeling rotating exotic objects \cite{Smailagic:2010nv,Bambi:2013ufa, Ghosh:2014pba,Dymnikova:2015hka,Ghosh:2015ovj,Atamurotov:2015xfa, Lamy:2018zvj,Sakti:2019iku,PhysRevD.100.024028}. Recently, we examined properties of the energy-momentum tensor of these Gurses-Gursey rotating systems \cite{Beltracchi2021a}, in particular, whether they satisfy an equation of state. The Newman-Janis algorithm was further extended by Drake and Szekeres to create rotating spacetimes from general static spherically symmetric metrics \cite{Drake:1998gf}, the purpose of this paper is to examine the energy-momentum tensors of these more general Drake-Szekeres rotating systems.

The Kerr and Kerr-Newman metrics have a clear interpretation of rotating and rotating charged black holes. Other systems with more physical properties than black holes can be analyzed in more detail to examine the correspondence between the rotating and nonrotating versions. We expect several connections between the rotating and nonrotating systems if they result from the same physical substance. We expect the Segre type of the nonrotating system is a specialization of the rotating system, for instance, a system that allows the pressures to be different along different axes in the rotating solution may be in a degenerate state with isotropic pressures in the static spherically symmetric solution. Also, the rotation should cause momentum density terms which can be undone locally with an appropriate comoving boost. Finally, we would expect that any relation between energy, pressure, and stress obeyed by the underlying physical substance can be satisfied in both the rotating and nonrotating stress-energy tensors.

The Newman-Janis algorithm is not specifically designed with the preservation of physical properties in mind and is known to distort the physical behavior in certain situations, such as not producing properly rotating monopole fields for Born-Infeld electrodynamics sources \cite{Lombardo_2004}. We found that for a specific class of Kerr-Schild systems could be interpreted as consisting of a substance obeying an equation of state in both their Gurses-Gursey rotating and nonrotating forms \cite{Beltracchi2021a}, but for general Gurses-Gursey systems, the equation of state obeyed by the spherical system is not obeyed by the rotating system. The Drake-Szekeres systems are substantially more complicated, although in their paper Drake and Szekeres find the only perfect fluid system generated by their method is the vacuum Kerr solution \cite{Drake:1998gf}, so even if one had a perfect fluid system originally the Drake-Szekeres rotating version will not be a perfect fluid. In this paper, we manage to reproduce this perfect fluid/Kerr result of Drake-Szekeres. We also find that in general, the Segre type of Drake-Szekeres rotating systems can change in rather unusual ways, such as generating regions of Segre type $[11Z\bar{Z}]$ in rotating systems, where the nonrotating versions are $[(111,1)]$ everywhere. This behavior makes physical interpretation of the Drake-Szekeres systems in general difficult.

\section{The Newman-Janis method for general systems}
We refer to Part I~\cite{Beltracchi2021a} for a review of the original Newman-Janis method and its extension to Kerr-Schild systems by Gurses and Gursey~\cite{Gurses:1975vu, Lamy:2018zvj}. Here we focus on the generalization of the method to work on arbitrary static spherically
 symmetric spacetimes due to Drake and Szekeres~\cite{Drake:1998gf}.
 Similarly to what we did in Part I, we may examine the problem in terms of correspondance between a spherical metric in Schwarzschild coordinates and a rotating metric in Boyer-Lindquist coordinates. 
 
(1) We write the metric of the general static spherically symmetric spacetime in the form
\begin{subequations}
\label{gensphere}
\begin{align}
ds^2 = - f(r) \, dt^2 + \frac{dr^2}{h(r)} + r^2 \, d\theta^2 + r^2 \sin^2\theta \, d\phi^2 ,
\end{align}
with
\begin{align}
h(r) = 1 - \frac{2m(r)}{r} , \qquad f(r) =  \frac{h(r)}{[ j(r)]^2}, 
\end{align}
\end{subequations}
and $j(r)$ defined in such a way that it is positive and real.
The $dt^2$ term is timelike when $h(r)>0$ and spacelike when $h(r)<0$. The signature of the spacetime metric imposes $f(r) h(r) > 0$. 

(2) We obtain the corresponding rotating metric $ds^2 = g_{\mu\nu} \, dx^\mu \, dx^\nu$ as\footnote{ The notation here is rather unfortunate. The functions $h$ and $j$ in Drake-Szekeres are not the same functions we use, as our definition of $j$ is from \cite{Hartle:1967he}. The easiest basis of comparison is their Eq. (22) with our Eq.~(\ref{DSNJmet}) with the identification $\chi=\cos \theta$ and the replacement of their $j(r)$ and $k(r)$ with our $r^2 h(r)$ and $r^2 j(r)$, respectively, and taking  $a\rightarrow0$ to recover our Eq.~(\ref{gensphere}). Additionally, they use the mostly minus metric convention, and they write $f=e^{2\Phi}$ while using complex coordinates, which makes it unclear if they assume $f>0$ or not.} 
\begin{align}
    ds^2=-\Sigma\frac{a^2 \cos^2\theta+r^2 h}{\Sigma_j^2}\, dt^2+2  \sin^2\theta\frac{(h-j)r^2 a \Sigma}{\Sigma_j^2} \, dt \, d\phi+\frac{\Sigma}{\Delta}\, dr^2+\Sigma \, d\theta^2+\nonumber\\
    \Sigma \sin^2\theta \frac{(a^2+jr^2)^2-a^2 \Delta \sin^2\theta }{\Sigma_j^2} \, d\phi^2 ,
    \label{DSNJmet}
\end{align}
where
\begin{subequations}
\begin{align}
& \Sigma = r^2 + a^2 \cos^2 \theta ,
\\
& \Delta = r^2 h(r) + a^2 = r^2 - 2 r m(r) + a^2 ,
\\
& \Sigma_j = r^2 j(r) + a^2 \cos^2 \theta.
\end{align}
\end{subequations}
It is useful to rearrange the terms in the metric to show a set of four mutually orthogonal one-forms,
\begin{align}
    ds^2=-\frac{\Sigma\Delta}{\Sigma_j^2}\, (dt-a \sin^2 \theta\, d\phi)^2+\frac{\Sigma}{\Delta}\, dr^2+\Sigma \, d\theta^2+\frac{\Sigma\sin^2\theta }{\Sigma_j^2}\, \big[(a^2+jr^2)d\phi-a \, dt \big]^2.
    \label{DSforms}
\end{align}

We will refer to physical systems with metric as in Eqs.~\eqref{DSNJmet} or~\eqref{DSforms} as Drake-Szekeres rotating systems, with corresponding nonrotating metric in Eq.~\eqref{gensphere}.

\section{Static spherical systems}
  
  The energy-momentum tensor for the general static spherically symmetric seed metric given by Eq.~(\ref{gensphere}) is
 \begin{subequations}
   \label{spT}
     \begin{align}
T^t_{~t} = - \rho, \qquad T^r_{~r} = p_r , \qquad T^\theta_{~\theta} = T^\phi_{~\phi} = p_T ,
\label{sphereThl}
\end{align}
with
\begin{align}
 &     	\rho=\frac{m'}{4\pi r^2}, \\
 & 	p_r=-\frac{m'}{4\pi r^2}-\frac{(r-2m) j' }{4 \pi  r^2 j } \label{prsph1}\\
&      p_T=- \frac{m''}{8\pi r} + \frac{(3rm'-r-m)j'}{8\pi r^2 j} + \frac{(r-2m)(j')^2}{4\pi r j^2} - \frac{(r-2m) j''}{8\pi r j}
   \label{sphtmv}
  \end{align}
 \end{subequations}
 Recall that $h$ and $j$ are functions of $r$ only. An alternate form of Eq.~\eqref{sphtmv} is
\begin{align}
&     p_r'=- (\rho+p_r) \frac{f'}{f}+ \frac{2 (p_T-p_r)}{r} .
\label{eq:anisotropic_TOV}
 \end{align}
%\end{subequations}
The latter equation is the conservation of energy equation $\del_\mu T^{\mu\nu}=0$ and is the anisotropic form of the Tolmann-Oppenheimer-Volkov equation~\cite{1974ApJ...188..657B}.

Often the equations of state of the physical system are given \textit{a priori}, and then added to the differential equations~\eqref{spT} to result in a set of equations which is solved for $h(r)$ and $j(r)$. Alternately, one may start with given $j(r)$ and $h(r)$, compute the energy density and pressures by means of Eqs.~(\ref{spT}), and analyze any relation between them. 

In terms of Segre types, Eq.~(\ref{spT}) shows that the energy-momentum tensor is generally of Segre type [(11)1,1], with possible  degenerate cases [(111),1],  [1(11,1)], [(11)(1,1)] and [(111,1)]. In particular, it is never of Segre types containing  [$Z\bar{Z}$] or $[2]$. Information about Segre types and notation may be found in \cite{Stephani:2003tm}, but as a brief overview it denotes the eigenvalue and eigenvector structure of the Ricci or energy-momentum tensors with mixed indices. Eigenvalue degeneracy is denoted by grouping in parentheses, the eigenvalue associated with a timelike eigenvector (if present) comes after a comma, complex conjugate pairs of eigenvalues are denoted with $Z\bar{Z}$, and systems with double or triple null eigenvectors are denoted with $2$ or $3$ rather than the $1$s. The [(111),1] case (three equal ``space" eigenvalues and a distinct ``time" eigenvalue) is the case of a perfect fluid, and the [(11)(1,1)] and [(111,1)] cases are Kerr-Schild spacetimes and were discussed in Part I.
 
 If $j ={\rm const}$, it is easy to see from Eq.~(\ref{prsph1}) that $-\rho=p_r$ and the energy-momentum tensor becomes of Segre type [(11)(1,1)]. It is important to note that the components $T^\mu_{~\nu}$ in~\eqref{spT} assume the same value whatever constant value $j$ has, because all the $j$-dependent terms in them are proportional to either $j'$ or $j''$.  A metric of the form~(\ref{gensphere}) with $j={\rm const}$ is related to the spherical Kerr-Schild metric of the same form, which has $j=1$ and $f(r)=h(r)$, by the coordinate transformation $dt^*=dt/j$, and therefore describes the same manifold. We will see later, however, that the rotating solution provided by the Drake-Szekeres method is different for different constant values of $j$.

 If the pressure is isotropic ($p_T=p_r$), the energy-momentum tensor is of Segre type [(111),1], which is the Segre type of a perfect fluid.  This occurs when
 \begin{align}
&   (2 h-r^2 h'' -2) j^2 + 3 r^2 j  j' h' - 4 r^2 h  (j')^2+2 r^2 h j  j'' = 2 r h j j'. \label{eq:pT=pr}
\intertext{It is also possible to have Segre type [1(11,1)] when $p_T=-\rho\ne p_r$. This occurs when}
&   (2 h-r^2 h'' -2) j^2 + 3 r^2 j  j' h' - 4 r^2 h  (j')^2+2 r^2 h j  j'' = - 2 r h j j'.
\end{align}
Finally, a vacuum Segre type [(111,1)], for which $-\rho=p_r=p_T$, occurs for $j={\rm const}$ (from Eq.~\eqref{prsph1} since $h$ is not identically zero) and $2 h -r^2 h'' -2=0$ (from Eq.~\eqref{eq:pT=pr} with $j'=j''=0$). Substituting $h$ for $m$ in the latter equation and solving gives the Schwarzschild/de Sitter mass function $m(r)= M + \Lambda r^3/6$. This is thus the only case of vacuum Segre type [(111,1)].

 Type [(11)1,1] has three distinct eigenvalues so the system cannot fully be described by an equation of state between just two of them, and the examples in the literature have been defined in various ways. Detailed calculations of these sorts of systems date back to Bowers and Liang \cite{1974ApJ...188..657B}, although much of the essential physics was worked out by Lemaitre \cite{1997GReGr..29..641L}. Another relevant example is anisotropic ``gravastar" models which assign $\rho(r)$ and $p_r(r)$ \cite{eosgravastar}, or $\rho(r)$ and $p_r(\rho)$ \cite{chirenti2007tell}, and use the equations of~(\ref{spT}) to compute $h$ and $j$ (or some other notation for the $g_{tt}$ and $g_{rr}$ metric functions) and $p_T$.

 For types [(111),1] and [1(11,1)] there are only two distinct eigenvalues. Similarly to the type [(11)(1,1)] case in Sec. II, if either eigenvalue is an invertible function of $r$ then one may use that to find an equation of state between the two eigenvalues. Usage of an equation of state for perfect fluid type [(111),1] is especially common, and there are several examples of the perfect fluid in the literature. Using the condition that $p_T=p_r=p$ with Eq.~(\ref{eq:anisotropic_TOV}) gives the Tolman-Oppenheimer-Volkoff equation, which may be solved given an equation of state (at least numerically). There are also analytic exact solutions in this class. For instance, the Schwarzschild star has
 \begin{subequations}
  \begin{align}
&      h=\left(1-\frac{2M r^2}{R^3}\right) , \\
 &     f=\frac{1}{4}\left(3\sqrt{1-\frac{2M}{R}}-\sqrt{1-\frac{2M r^2}{R^3}}\right)^2.
  \end{align}
 \end{subequations}
  The equation of state for this (in addition to the $p_\perp=p_r=p$ isotropy condition) is 
  \begin{equation}
      \rho(p)={\rm const}=\frac{3M}{4 \pi R^3}.
  \end{equation}
Another analytic perfect fluid solution is the relativistic isothermal sphere \cite{1969Afz.....5..223B,1970ApJ...160..875B,1972grec.conf..185C,Chavanis:2007kn}, which arises as the solution with the equation of state
\begin{align}
    p=k\rho.
\end{align}
The metric functions are
\begin{subequations}
 \begin{align}
    h=\frac{1+6k+k^2}{1+2k+k^2}\\
    f=(r(1+k))^{\frac{4k}{1+k}}.
\end{align}
\end{subequations}

In contrast to the relatively common perfect fluid [(111),1] solutions and Kerr-Schild [(11)(1,1)] solutions, we are unaware of any spherically symmetric solutions with Segre type [1(11,1)].
  
Segre type [(111,1)] is vacuum energy and can only have the equation of state $p=-\rho$. 

\section{Rotating energy-momentum tensors}

Drake-Szekeres rotating systems are stationary and axisymmetric, with Killing vectors given in $(t,r,\theta,\phi)$ coordinates by
\begin{align}
K^\mu = \delta^\mu_{t} , \qquad \tK^\mu = \delta^\mu_{\phi} .
\end{align}
The azimuthal Killing vector $\tK^\mu$ is normalized as usual with $0\le \phi \le 2 \pi$. The time-translation Killing vector $K^\mu$ is instead allowed to have a normalization different from the usual one, in which $K^\mu$ equals $\delta^\mu_{t}$ in the asymptotically flat region of large $r$ values, because in general a Drake-Szekeres system may not be asymptotically flat. Moreover, $K^\mu$ is timelike when $\Delta>0$ and spacelike when $\Delta<0$, so in our choice of metric signature
\begin{align}
\sd K^\mu K_\mu < 0 ,
\end{align}
where 
\begin{align}
\sd=\sign\Delta ,
\end{align}
i.e., $\sd=1$ for $\Delta>0$ and $\sd=-1$ for $\Delta<0$.

It is useful to introduce the projectors $P^\mu_{~\nu}$ onto the two-dimensional space spanned by the Killing vectors and $Q^\mu_{~\nu}$ onto its two-dimensional orthogonal complement,
\begin{subequations}
\label{eq:P_Q_projectors}
\begin{align}
P^\mu_{~\nu}  & = G^{KK} K^\mu K_\nu + G^{K\tK} K^\mu \tK_\nu + G^{\tK K} \tK^\mu K_\nu + G^{\tK\tK} \tK^\mu \tK_\nu ,
\label{eq:P_projector} \\
Q^\mu_{~\nu} & = \delta^\mu_{~\nu}  - P^\mu_{~\nu} .
\label{eq:Q_projector}
\end{align}
\end{subequations}
Here $G^{AB}$ is the inverse of the $2\times2$ Gram matrix $G_{AB}$ of the vectors $K$ and $\tK$ given by $G_{KK} = K^\alpha K_\alpha$ , $G_{K\tK}=G_{\tK K} = K^\alpha \tK_\alpha$, $G_{\tK\tK} = \tK^\alpha \tK_\alpha$. On the horizon the linear combination $K^\mu + \omega \tK^\mu$ is a null vector and the Gram matrix $G_{AB}$ becomes degenerate. We omit the horizon from this discussion.

Because of the separate invariance of the metric under $t \mapsto -t$ and $\phi \mapsto -\phi$, the components $T_{tr},T_{t\theta},T_{\phi r},T_{\phi \theta}$ must be 0, and the $T_{\mu\nu}$ splits into a $t\phi$ block and an $r\theta$ block, which correspond respectively to its projections $P^\mu_{~\alpha} P^\beta_{~\nu}T^\alpha_{~\beta} $ and $Q^\mu_{~\alpha} Q^\beta_{~\nu} T^\alpha_{~\beta} $ onto the space spanned by the Killing vectors and onto the space orthogonal to it. Thus the components $T_{tt},T_{\phi \phi},T_{t \phi}, T_{rr},T_{\theta\theta},T_{r \theta}$ may be nonzero.

  The components $T_{\mu\nu}$ themselves in the coordinate basis are rather lengthy. The form of the metric in Eq.~\eqref{DSforms} suggests the introduction of an orthonormal tetrad $e^\alpha_{~\hat{\alpha}}$ comprised of the normalized dual vectors of the one-forms in Eq.~\eqref{DSforms},
  \begin{align}
     e^\alpha_{~\hat{\alpha}}=\left(
\begin{array}{cccc}
 \frac{a^2+jr^2}{\sqrt{|\Delta|  \Sigma }} & 0 & 0 &
   \frac{a \sin \theta}{\sqrt{\Sigma}} \\
 0 & \sqrt{\frac{|\Delta| }{\Sigma }} & 0 & 0 \\
 0 & 0 & \sqrt{\frac{1}{\Sigma }} & 0 \\
 \frac{a}{\sqrt{|\Delta|  \Sigma }} & 0 & 0 &
   \frac{1}{\sin \theta \sqrt{\Sigma} } \\
\end{array}
\right) .
\label{tet2}
 \end{align}
Here the spacetime index $\alpha$ labels the rows and the orthonormal index $\hat{\alpha}$ labels the columns, and we assume $\Delta\ne0$.  The metric in this orthornormal frame can be computed as
\begin{align}
g_{\hat{\alpha}\hat{\beta}} = g_{\alpha\beta} e^\alpha_{~\hat{\alpha}} e^\beta_{~\hat{\beta}} = \diag( - \sd, \ \sd, \ 1, \ 1),
\end{align}
If $\Delta$ is negative, then the $r$ coordinate is timelike and the orthonormal metric is $g_{\hat{\alpha}\hat{\beta}}=\diag(1,-1,1,1)$ rather than the usual $g_{\hat{\alpha}\hat{\beta}}=\diag(-1,1,1,1)$ which applies when $\Delta>0$. 

The tetrad~\eqref{tet2} allows for considerable simplification of the orthonormal components $T_{\hat{\mu}\hat{\nu}}$ of the stress-energy tensor for the Drake-Szekeres system, but the tetrad~\eqref{tet2} is not the principal frame of the stress-energy tensor where it is diagonal. The components in $\hat{0},\hat{1},\hat{2},\hat{3}$ follow the pattern
 \begin{equation}
    T_{\hat{\mu}\hat{\nu}}=\left(\begin{array}{cccc}
        - \sd \muo &0 & 0& \sd \siiio \\
        0 & \sd \mui&\sIii &0\\
        0&\sIii&\muii&0\\
        \sd \siiio&0&0&\muiii
    \end{array}\right) ,
    \label{simpleT}
\end{equation}
with
\begin{align}
\sIii = \muIii  \sqrt{|\Delta|}  \sin\theta, 
\qquad
\siiio = \muiiio  \sqrt{|\Delta|}  \sin\theta, 
\end{align}
where the quantities $\hat{\mu}_i(r,\theta)$ and $\hat{\mu}_{ij}(r,\theta)$, besides being rational functions in $\cos\theta$, are polynomials in $\Delta$ and its derivatives $\Delta'$ and $\Delta''$, or equivalently $m$ and its derivatives $m'$ and $m''$. They 
are listed in the Appendix, together with the expression of the coordinate components $T_{\mu\nu}$ and $T^{\mu}_{~\nu}$ in terms of the $\hat{\mu}$'s.

For $j=1$, one has $\siiio = \sIii = 0$, and the orthonormal frame diagonalizes the energy-momentum tensor. This is the Kerr-Schild case discussed in Part I~\cite{Beltracchi2021a}, where we found%
\begin{subequations}
\label{eq:spinDENP}
\begin{align}
&    -\rho=p_\parallel=\muo=\mui=-\frac{r^2 m'}{4 \pi \Sigma^2} , \\
&    p_\perp = \muii=\muiii=- \frac{rm''}{8\pi \Sigma} - \frac{a^2 \cos ^2\theta \, m'}{4 \pi  \Sigma ^2} .
%&    \siiio =\sIii=0 .
\end{align}
\end{subequations}
In this case, the only possible Segre types are [(11)(1,1)] and [(111,1)], which were discussed in Part I~\cite{Beltracchi2021a}, where it was concluded that an equation of state between the distinct eigenvalues that applies to both rotating and nonrotating systems exists only for a special family of Kerr-Schild spacetimes that includes the Kerr and Kerr-Newman black holes, as well as rotating spacetimes whose mass function in the nonrotating limit contains a constrained superposition of a cloud of strings term, a Reissner-Nordstrom term, a cosmological constant term, and a Schwarzschild term. 

\subsection{Segre types}

The Segre type is found from the eigenvalues of $T^\mu_{~\nu}$ (the energy-momentum, Ricci and Einstein tensors with mixed indices have the same Segre type, determined by their traceless parts). 
 Because the metric in the orthonormal frame is simply $g_{\hat{\mu}\hat{\nu}} = \diag(-\sd,\sd,1,1)$, we get
\begin{equation}
    T^{\hat{\mu}}_{\ \,\hat{\nu}}=\left(\begin{array}{cccc}
        \muo &0 & 0& - \siiio \\
        0 &  \mui&\sd \sIii &0\\
        0&\sIii&\muii&0\\
        \sd \siiio&0&0&\muiii
    \end{array}\right).
\end{equation}
The block structure allows us to consider the eigenvalues of the $t\phi$ and $r\theta$ blocks separately. The characteristic equations of the two blocks are
\begin{subequations}
\begin{align}
(\lambda - \muo)(\lambda - \muiii) + \sd \siiio^2 = 0 , \label{eq:tphi_char} \\
(\lambda - \mui)(\lambda - \muii) - \sd \sIii^2 = 0 , \label{eq:rtheta_char} 
\end{align}
\end{subequations}
with discriminants 
\begin{subequations}
\label{eq:DD}
\begin{align}
\DIii & = (\mui- \muii)^2+4\,\sd \sIii^2, \\
\Diiio & = (\muiii-\muo)^2-4\, \sd \siiio^2 ,
\end{align}
\end{subequations}
respectively. Despite the appearance of $\sd$ in these formulas, the discriminants $\DIii$ and $\Diiio$ are polynomials in $\Delta$, because $\sd \siiio^2=\sin^2\theta\,\muIii^2 \Delta$ and $\sd \sIii^2 =\sin^2\theta\,\muiiio^2\Delta$.
The eigenvalues follow as the solutions
\begin{subequations}
\label{eq:lambda_rtheta_tphi}
 \begin{align}
&    \lambda_{r\theta}^{\pm}=\frac{1}{2} \Big(\BIii\pm\sqrt{\DIii} \, \Big), \label{eq:lambda_rtheta} \\
 &   \lambda_{t\phi}^{\pm}=\frac{1}{2} \Big(\Biiio\pm\sqrt{\Diiio} \, \Big). \label{eq:lambda_tphi}
\end{align}
\end{subequations}
where
\begin{align}
\BIii = \mui+\muii, \qquad \Biiio = \muiii+\muo
\label{eq:BB}
\end{align}
are the traces of the $r\theta$ and $t\phi$ blocks respectively.

The traces $\BIii$ and $\Biiio$, and the discriminants $\DIii$ and $\Diiio$, can also be written in an invariant form by means of the projectors $P^\mu_{~\nu}$ and $Q^\mu_{~\nu}$ in Eqs.~\eqref{eq:P_projector} and~\eqref{eq:Q_projector}, after recalling that for a $2\times2$ matrix the trace equals the sum of its eigenvalues $\lambda_++\lambda_-$ and the discriminant equals the square of their difference $(\lambda_+-\lambda_-)^2$. Thus, using $(\lambda_++\lambda_-)^2+(\lambda_+-\lambda_-)^2 = 2 (\lambda_+^2 + \lambda_-^2) $ and the matrix notation $T=(T^\mu_{~\nu})$, $P=(P^\mu_{~\nu})$, $Q=(Q^\mu_{~\nu})$,
\begin{subequations}
\label{eq:BD_covariant}
\begin{align}
& \Biiio = \tr(PT) = P^\mu_{~\nu} T^{\nu}_{~\mu} , \\
& \Diiio = 2\tr(PTPT) - [\tr(PT)]^2 = 2 P^\mu_{~\nu} T^{\nu}_{~\alpha} P^\alpha_{~\beta} T^{\beta}_{~\mu} - (P^\mu_{~\nu} T^{\nu}_{~\mu} )^2 , \\
& \BIii = \tr(QT) = Q^\mu_{~\nu} T^{\nu}_{~\mu} , \\
& \DIii= 2\tr(QTQT) - [\tr(QT)]^2 = 2 Q^\mu_{~\nu} T^{\nu}_{~\alpha} Q^\alpha_{~\beta} T^{\beta}_{~\mu} - (Q^\mu_{~\nu} T^{\nu}_{~\mu} )^2 .
\end{align}
\end{subequations}

For $\Delta>0$, $\DIii\ge0$ while $\Diiio$ can be positive, negative, or zero. Thus the $r\theta$ block has Segre type [11] or [(11)], the latter occurring for $\DIii=0$, and the $t\phi$ block has Segre type [1,1], [(1,1)], [2], or [\zzbar] according to the conditions $\Diiio>0$, $\Diiio=0$ with $\siiio=0$, $\Diiio=0$ with $\siiio\ne0$, or $\Diiio<0$, respectively.  For $\Delta<0$, the roles of the $r\theta$ and $t\phi$ blocks are interchanged, and their Segre types can be found by exchanging $\Diiio \leftrightarrow \DIii$, $\siiio \leftrightarrow \sIii$ in the previous sentences.

One expression of the eigenvectors in $(\hat{0},\hat{1},\hat{2},\hat{3})$ for the $[11]$, $[1,\!1]$, and $[Z\bar{Z}]$ cases in the $t\phi$ and $r\theta$ blocks is
\begin{subequations}
\label{DSeigenvactors}
 \begin{align}
    V_1^{\hat{\mu}}&=(- \siiio, \ 0, \ 0, \ \lambda_{t\phi}^+ -\muo)\\
    V_2^{\hat{\mu}}&=(\lambda_{t\phi}^--\muiii, \ 0, \ 0, \ \sd \siiio)\\
    V_3^{\hat{\mu}}&=(0, \ \sd \sIii, \ \lambda_{r\theta}^+- \mui, \ 0)\\
    V_4^{\hat{\mu}}&=(0, \ \lambda_{r\theta}^--\muii, \ \sIii, \ 0).
\end{align}
\end{subequations}
These eigenvectors  become $0$ vectors if the respective block is [(11)] or [(1,1)], but for these degenerate Segre types any vector of the form $P^\mu_{\ \nu}V^\nu$ or $Q^\mu_{\ \nu} V^\nu$ will be an eigenvector in the degenerate subspace. A $2\times2$ block of Segre type [2] is a defective matrix and has only one proper eigenvector, for instance here $V_1\rightarrow sign(\siiio)V_2$ when the $t\phi$ block is type [2]. 

Notice when $\Delta>0$, the $\DIii$ term inside the square root of $\lambda_{r\theta}$ [Eq.~(\ref{eq:lambda_rtheta})] is the sum of two squares, as is the $\Diiio$ term from the square root in $\lambda_{t\phi}$ [Eq.~(\ref{eq:lambda_tphi})]  when $\Delta<0$ . This means that there can be no complex eigenvalues and the Segre type must be [11] or [(11)] for the $t\phi$ block when $\Delta<0$ and for the $r\theta$ block when $\Delta>0$.
Another important property of the eigenvalues in the $r\theta$ block is that when $\Delta>0$, they can only be degenerate if $\sIii=0$ and $\mui=\muii$. The observation of $\sIii=0$ being necessary for degenerate eigenvalues in this block\footnote{If $\Delta<0$ everywhere an analogous argument can be made about the $t\phi$ block.} allows us to verify the assertion from \cite{Drake:1998gf} that the only Drake-Szekeres system with a perfect fluid energy-momentum tensor is the Kerr solution. We see from Eq.~(\ref{sigmainj}) that the only way to have $\sIii=0$ for all $r,\theta$ is $j=1$, or a standard Newman-Janis system of the Kerr-Schild type (covered in Part I~\cite{Beltracchi2021a}). For a perfect fluid, all the spacelike eigenvalues must be the same, so by the eigenvalues from Eqs.~(\ref{eq:spinDENP}) must be equal. This in turn implies $m=M$, which is the Kerr solution.
  
In the instance where $\Delta>0$, the $r\theta$ block can only be Segre [11] or possibly [(11)], but the $t\phi$ block has more possible Segre types.
The Segre type of the $t\phi$ block is controlled by the discriminant $\Diiio$. If $\Diiio>0$, $\siiio\ne0$, and $\Delta>0$, then the Segre type of the $t\phi$ block is [1,1]. In this case we may diagonalize the $t\phi$ block with a local azimuthal boost of velocity
\begin{align}
    \beta=\sign(\muo-\muiii)\,\frac{|\muo-\muiii|-\sqrt{\Diiio} }{2 \siiio} .
\end{align}
We can ascribe this to a rotation of the energy comoving frame at azimuthal velocity $\beta$ with respect to the local inertial frame~\eqref{tet2}.

If $\Diiio=0$, there are two possibilities, [(1,1)] and [2], for the Segre type of the $t\phi$ block. If $ \siiio =0$ and $\muiii=\muo$, then it is vacuum energy like and [(1,1)]. If instead $ \siiio \ne0$, it is type [2].
If $\Diiio<0$, then the $t\phi$ block is Segre type [$Z\bar{Z}$]. Such an energy-momentum tensor is less physically relevant, but we have determined it may occur for certain Drake-Szekeres systems. 

If instead $\Delta<0$, then the $t\phi$ block must be Segre [11] or [(11)] and the $r\theta$ block has the extended possibilities.  The discriminant $\DIii$ controls the Segre type of the $r\theta$ block when $\Delta<0$. If $\DIii>0$, the Segre type is [11]. If $\DIii=0$, the Segre type is [(1,1)] for $\sIii=0$ and [2] otherwise. If $\DIii<0$, the Segre type is [$Z\bar{Z}$]. We summarize the Segre type of the blocks in Table \ref{segretable}.

\begin{table}[]
\begin{center}
\begin{tabular}{ |c|c|c|c|c|c| }
 \hline
 \multicolumn{6}{|c|}{$\Delta>0$} \\
 \hline
 \multicolumn{4}{|c|}{$t\phi$ block} & \multicolumn{2}{|c|}{$r\theta$ block}\\
 \hline
 $\Diiio>0$& \multicolumn{2}{|c|}{$\Diiio=0$} & $\Diiio<0$& $\DIii=0$ &$\DIii\ne0$\\
 \hline
 &$ \siiio =0$&$ \siiio\ne0$& & &\\
 $[1,1]$&$[(1,1)]$&$[2]$&$[Z\bar{Z}]$&$[(11)]$&$[11]$\\
 \hline
\end{tabular}
\begin{tabular}{ |c|c|c|c|c|c| }
 \hline
 \multicolumn{6}{|c|}{$\Delta<0$} \\
 \hline
 \multicolumn{4}{|c|}{$r\theta$ block} & \multicolumn{2}{|c|}{$t\phi$ block}\\
 \hline
 $\DIii>0$& \multicolumn{2}{|c|}{$\DIii=0$} & $\DIii<0$& $\Diiio=0$ &$\Diiio\ne0$\\
 \hline
 &$\sIii=0$&$\sIii\ne0$& & &\\
 $[1,1]$&$[(1,1)]$&$[2]$&$[Z\bar{Z}]$&$[(11)]$&$[11]$\\
 \hline
\end{tabular}
\caption{Table showing the possible Segre types for the blocks. The Segre type of the energy-momentum tensor is obtained by juxtaposing the Segre types of the $t\phi$ and $r\theta$ blocks, with the addition of possible degeneracies between the combined eigenvalues. For example, $[2]$ in the $t\phi$ block and $[(11)]$ in the $r\theta$ block combine into the Segre type $[2(11)]$ or its degenerate case $[(211)]$. \label{segretable}}
\end{center}
\end{table}
In total, we find the Segre type for a Drake-Szekeres rotating metric may be quite general. We have for instance [(111,1)] for Kerr  ($m={\rm const}, j=1$), [(11)(1,1)] for ($j=1$). We explicitly show [211], [111,1], and $[11Z\bar{Z}]$ Segre types at particular points in an example we give below. Additional degeneracies may be possible at particular spacetime points, i.e. the Segre type cannot globally be perfect fluid [(111),1] for nonzero $a$, but this could occur on some subspace. The structure of the eigenvectors in Eq.~(\ref{DSeigenvactors}) does, however, forbid Segre type [31] and its degenerate case $[(31)]$. 

To summarize our Segre type analysis, the Drake-Szekeres generalization of the Newman-Janis algorithm can produce rotating system with points of any Segre type but $[31]$ and $[(31)]$.

\section{Simple example: ``spinning Minkowski"}
\label{sec:example}
 One spacetime simple enough to  handle explicitly and see various possible Segre behaviors is the rotating Newman-Janis version of the spherical space with $m(r)=0$, $j(r)={\rm const}$, which is Minkowski space with a scaled time coordinate.  This seed spacetime is the most general spherically symmetric static spacetime that has no matter content ($T_{\mu\nu}=0$), zero curvature, no singularities, and importantly no event horizons (so $\Delta>0$ and the $r$ coordinate is spacelike). A Newman-Janis rotating system based on de Sitter space is referred to as ``rotating de Sitter" in various papers ~\cite{Ibohal:2004kk,2006PhLB..639..368D, deUrreta:2015nla}. We could likewise call metric \eqref{DSNJmet} with $m=0$, $j={\rm const}$ a ``spinning Minkowski" space. These names should not be taken too seriously however, as the ``rotating de Sitter" does not usually have constant curvature, and the ``spinning Minkowski" is not generally flat. 
  However, we find all 20 independent degrees of freedom in the Riemann tensor have at least one power of $a$ and one power of $j-1$ (or are identically 0) showing that if $a=0$ or if $j=1$ we do in fact have flat space. We omit a list of all components of the Riemann tensor for the sake of space, but for reference the Ricci scalar is
\begin{align}
    \mathcal{R}=\frac{a^2 (j-1)}{\Sigma ^2
   \Sigma_j^2} \Big[ 4 (r^2+a^2\chi^2) (2r^2\chi^2+a^2\chi^2-r^2)
   +2 (j-1) r^2 (5r^2\chi^2-5a^2\chi^2-2r^2-a^2\chi^4) \Big]
\end{align}
and the Kretchmann scalar is
\begin{align}
  &\mathcal{K}= \frac{4 a^2 (j-1)^2}{\Sigma^4\Sigma_j^4}\Big\{
   a^6 r^4 \chi ^4 \left[
     (11 j^2+2j+19) \chi ^4+4 (4 j^2-9j-4) \chi^2+9 j^2+10 j+9
   \right]+
   \nonumber\\&
   a^2 r^8 \left[
     (19 j^2+2j+11) \chi ^4+2 j (9j+8) \chi^2+2j^2
   \right]
   -2 a^8 r^2 \chi ^6 \left[
   5\chi ^4+(8 j+9) \chi ^2+4 (j-3) 
   \right]+
   \nonumber\\&
   4 a^4 r^6 \chi ^2 \left[
     (5 j^2+11j+5) \chi ^4+(4j^2+9j-4) \chi ^2+2 j (3 j-1)
   \right]
   +2 a^{10} \chi^8 (\chi^2+1)
   -2 j^2 r^{10} (5 \chi ^2+1)
   \Big\}
\end{align}

\subsection{Block Segre types}
 The Drake-Szekeres ``spinning Minkowski" spacetime results in the following expressions for the invariants  $\BIii$, $\Biiio$, $\DIii$, and $\Diiio$, 
\begin{subequations}
\label{eq:BDBD}
\begin{align}
    \BIii&=\frac{a^2(j-1)}{4\pi\Sigma ^2 \Sigma_j^2} \Big[ -a^2 (4 r^2+a^2) \chi^4 - r^2 (4 j r^2-3 j a^2 + 3 a^2 ) \chi^2  +j
   r^4 \Big] ,
\\
   \DIii &= \frac{a^4 (j-1)^2}{16\pi^2 \Sigma ^4 \Sigma_j^4} \Big[
   a^4 \big[(j-1)(j-13)r^4-2(j+11)a^2r^2+a^4\big]\chi^8
   \nonumber \\ & \qquad
   + 2 a^2 r^2 \big[ (j-1)^2r^4+(2j^2-7j-7)a^2r^2-2(j-5)a^4 \big] \chi^6
   \nonumber \\ & \qquad
   +r^4 \big[ (j-1)(13j-1)r^4 + 2 (7j^2+7j-2)a^2r^2 + 2(2j^2+11j+2)a^4 \big] \chi^4
    \nonumber \\ & \qquad
   +2jr^6\big[(11j+1)r^2+2(5j-1)a^2\big]\chi^2
   +j^2r^8
   \Big] ,
\\
\Biiio&=\frac{a^2 (j-1)}{4\pi\Sigma ^2 \Sigma_j^2} \Big[ a^2 (jr^2-r^2-a^2) \chi^4 - (j-1) r^2 (r^2-2a^2) \chi ^2 + j r^4\Big] ,
\\
 \Diiio&=  \frac{a^2 (j-1)^2}{16\pi^2 \Sigma ^3 \Sigma_j^4}\Bigg[ 
   a^6 (r^2+a^2) \chi^8 
   +a^4 r^2 (4 j^2 r^2-18 j r^2+13 r^2-6 j a^2+4 a^2 ) \chi^6 \nonumber\\&
   +a^2 r^4 (13 j^2 r^2-18 j r^2+4r^2+5 j^2 a^2 -5 a^2) \chi^4 
   +j r^6 (j r^2-4 j a^2+6 a^2 ) \chi^2 
   -j^2 r^8 \Bigg] . \label{Dmink}
\end{align}
\end{subequations}
The $t\phi$ eigenvalues~\eqref{eq:lambda_tphi}
are complex when $\Diiio<0$, giving a Segre type of [11$Z\Bar{Z}$] or possibly [(11)$Z\Bar{Z}$] if the $r\theta$ eigenvalues~\eqref{eq:lambda_rtheta}
happen to be degenerate, which occurs for $\DIii=0$. One can see from these expressions that when $a=0$ (no rotation) or $j=1$ (no time scaling in seed system), we see that the invariants go to zero as we would expect for a vacuum spacetime. 
Note that the polynomial term in square brackets in Eq.~\eqref{Dmink} determines the sign of $\Diiio$ and hence the Segre type, as the prefactor is always positive. At $\chi^2=0$, this polynomial becomes $-j^2r^8$, which is negative, so the Segre type at points on the equator $\cos\theta=0$ is [11$Z\Bar{Z}$] or [(11)$Z\Bar{Z}$]. At $\chi^2=1$, the polynomial in square brackets becomes $\left(a^2+r^2\right) \big[a^3+a (2-3 j) r^2\big]^2$, which is nonnegative for nonzero $a$, so there must be at least one root at each $r$ in the interval where $\Diiio=0$. Since the term is a polynomial in $\chi$, certain $r$ may have multiple roots. The Segre type of the $t\phi$ block  at the $\Diiio=0$ points will be [2] if $\siiio\ne 0$ and [(1,1)] if $ \siiio=0$ as well. 
The latter occurs when, see Eq.~\eqref{sigma30},
\begin{equation}
   \sin\theta\,  (j-1) (r^2-a^2\cos^2\theta) \, (jr^2-a^2\cos^2\theta)=0.
   \label{eq:sigma30=0}
\end{equation}
The solutions of~\eqref{eq:sigma30=0} are $r^2=a^2\cos^2\theta$, $r^2=a^2\cos^2\theta/j$, $j=1$, and $\sin{\theta}=0$. In order to have the [(1,1)] in the $t\phi$ block we simultaneously need $\Diiio=0$, which happens for $j=1$, which is an everywhere flat spacetime, and for  $\sin\theta=0$ with
\begin{equation}
    r=a/(3j-2)^{1/2}, \qquad \sin\theta=0,
    \label{(1,1)}
\end{equation}
which gives isolated points for $j>2/3$ where the Segre type of the $t\phi$ block is [(1,1)]. The other conditions $r^2=a^2\cos^2\theta$, $r^2=a^2\cos^2\theta/j$ do not have $\Diiio=0$ unless $j=1$ which is already covered.

\begin{figure}
    \centering
    \includegraphics{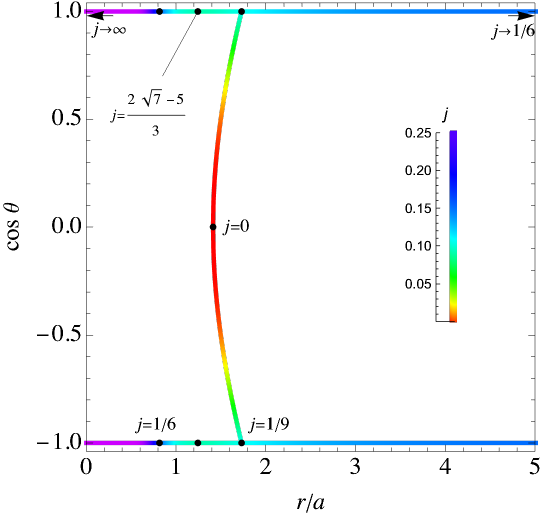}
    \caption{Points from Eq.~\eqref{(11)a}, the almost vertical arcline, and Eq.~\eqref{(11)b}, the lines at $\cos\theta=\pm1$, where the Segre type of the $r\theta$ block is [(11)] in the spacetime example in Sec.~\ref{sec:example}.}
    \label{rainbowj}
\end{figure}

The $r\theta$ block in this example has Segre type [11] when $\DIii\ne0$ and Segre type [(11)] when $\DIii=0$. The latter condition is a quadratic equation in $j$ which has real solutions only when its discriminant is greater or equal to zero,
\begin{align}
144 z^6 (1+z^2) \chi^6 (\chi^2-1)(\chi^2+z^2)^2(\chi^2+2-z^2)^2 \ge 0 ,
\end{align}
where $z=r/a$. Since $|\chi|\le 1$ and $z\ge0$, this discriminant cannot be greater than zero. It is zero when either $\chi^2=z^2-2$ or $\chi^2=1$. The former requires $\sqrt{2} \le z \le \sqrt{3}$, since $0\le\chi^2\le1$, and gives one point 
\begin{equation}
    r=a\sqrt{\frac{2}{1-\sqrt{j}}}, \qquad \cos\theta = \sqrt{\frac{2\sqrt{j}}{1-\sqrt{j}}} 
    \label{(11)a}  
\end{equation}
for every $0\le j\le1/9$ as solution of $\DIii=0$.
For the other condition $\chi^2=1$, points on the axis where the $r\theta$ block Segre type is [(11)] have $r$ coordinate satisfying the equation
\begin{align}
(6j-1)z^4 + (3j+1) z^2 -1 = 0,
\label{(11)b}
\end{align}
which has two solutions for $(2\sqrt{7}-5)/3\le j<1/6$, one solution for $j\ge1/6$, and no solutions for $j<(2\sqrt{7}-5)/3$. We show the points of degeneracy in the $r\theta$ block and their corresponding $j$ values in Fig. \ref{rainbowj}.

\iffalse
 \begin{figure}[t]
     \centering
     \includegraphics[width=8cm]{smallj.pdf}
     \includegraphics[width=8cm]{largej.pdf}
     \caption{Plots showing Segre types for values $j=1/10$ on the left and $j=3$ on the right. The gray regions are where $\Diiio<0$ and the Segre type of the $t\phi$ block is $[Z\Bar{Z}]$. The white regions are where $\Diiio>0$ and the Segre type of the $t\phi$ block is [11]. The boundary between the two regions (shown as black) with $\Diiio=0$ are roots of the polynomial in Eq.~\eqref{Dmink} and generally have Segre type [2] for the $t\phi$ block. Note the lobe structure present at smaller radii which indicate the existence of more than one root of the polynomial in $\Diiio$. There are isolated points where $\Diiio=0=\siiio$ with $|\chi|=1$ and Eq.~\eqref{(1,1)} is satisfied, in which case the Segre type of the $t\phi$ block is [(1,1)]; such points are shown with solid blue. The Segre type of the $r\theta$ block is generally [11], isolated points where it is the degenerate [(11)] given by formulas~\eqref{(11)a},~\eqref{(11)b} are shown with hollow red circles. \textbf{Note: cross block degeneracy means these figures are obsolete}.
      }
     \label{Picture}
 \end{figure}
\fi
 By plotting the polynomials in square brackets in Eq.~(\ref{Dmink}) at a given $j$ as functions of $r/a$ and $\cos\theta$, which we show in Fig. \ref{crossblockb}, we can see the negative regions where the Segre type is  [11$Z\bar{Z}$], the positive regions where it is [111,1], and the boundary between these regions where the Segre type is [112]. We also mark the isolated points with extra degeneracies such as [(11)1,1] and [11(1,1)]. 
 
 \subsection{Cross block degeneracy}
The conditions for degeneracy within a block are fairly simple, being either $\Diiio=0$ or $\DIii=0$. It is possible for there to be a degeneracy between blocks as well if any of the $t\phi$ block eigenvalues $\lambda_{t\phi}^\pm$ equals one of the $r\theta$  block eigenvalues $\lambda_{r\theta}^\pm$. If $\epsilon_{t\phi}=\pm1$ and $\epsilon_{r\theta}=\pm1$ correspond to $\lambda_{t\phi}^\pm$ and $\lambda_{r\theta}^{\pm}$, respectively, the relations~\eqref{eq:lambda_rtheta_tphi} transform the condition $ \lambda_{t\phi}^\pm = \lambda_{r\theta}^\pm$ into
\begin{align}
\Biiio-\BIii + \epsilon_{t\phi} \sqrt{\Diiio}- \epsilon_{r\theta} \sqrt{\DIii} = 0.
\label{crossblockdef}
\end{align}
Eliminating the square roots we obtain
\begin{equation}
    \big[(\Biiio-\BIii)^2-\Diiio-\DIii\big]^2-4\Diiio \DIii=0,\label{crossblockcond}
\end{equation}
regardless of which of the four possible combinations of $+$ and $-$ was chosen in \eqref{crossblockdef}. 

\begin{figure} [H]
     \centering
     \includegraphics[width=8cm]{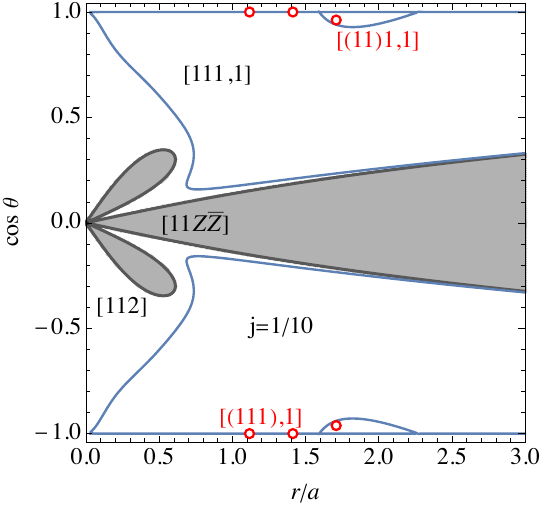}
     \includegraphics[width=8cm]{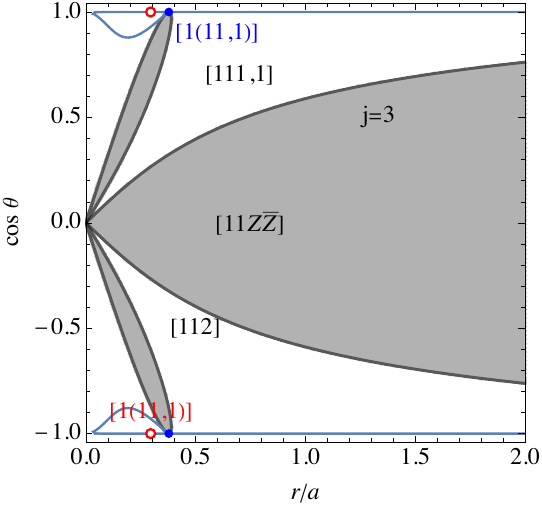}
     \caption{Plots showing Segre types for values $j=1/10$ on the left and $j=3$ on the right. The gray regions are where $\Diiio<0$ and the Segre type of the $t\phi$ block is $[Z\Bar{Z}]$. The white regions are where $\Diiio>0$ and the Segre type of the $t\phi$ block is [11]. The boundary between the two regions (shown as black) with $\Diiio=0$ are roots of the polynomial in Eq.~\eqref{Dmink} and generally have Segre type [2] for the $t\phi$ block. Note the lobe structure present at smaller radii which indicate the existence of more than one root of the polynomial in $\Diiio$. There are isolated points where $\Diiio=0=\siiio$ with $|\chi|=1$ and Eq.~\eqref{(1,1)} is satisfied, in which case the Segre type of the $t\phi$ block is [(1,1)]; such points are shown with solid blue. The Segre type of the $r\theta$ block is generally [11], isolated points where it is the degenerate [(11)] given by formulas~\eqref{(11)a},~\eqref{(11)b} are shown with hollow red circles. The blue line shows the location where \eqref{crossblockcond} is satisfied and there is a cross block degeneracy. Notice how it always stays outside the $[Z\bar{Z}]$ region. Due to the cross block degeneracy on the axis, some of the red and blue points (those satisfying both 5.11 and 5.8) have a triple degeneracy as in [(111),1] or [1(11,1)]. The [(111),1] are examples of perfect fluid behavior in the Drake-Szekeres system, but these are isolated occurrences rather than a perfect fluid full spacetime, which only occurs for the Kerr solution.}
     \label{crossblockb}
 \end{figure}

We can use the expressions in \eqref{eq:BDBD} in \eqref{crossblockcond} to find where the example $m=0,j={\rm const}$ spacetime has cross block degeneracies. In this case, \eqref{crossblockcond} is a polynomial in $j,\chi,r/a$. The first property worth mentioning is that when the $t\phi$ block has $[Z\bar{Z}]$ type, a cross block degeneracy is impossible because the eigenvalues of the $r\theta$ block must be real and can not be degenerate with a complex eigenvalue from the other block. In terms of \eqref{crossblockcond}, this can be seen in that the first term is a positive perfect square and the product $\Diiio \DIii$ of the $D$ terms is nonpositive since $\DIii\ge0$ and $\Diiio<0$ in the case of the  $[Z\bar{Z}]$. A second property is that when $\chi^2=1$, \eqref{crossblockcond} is satisfied independently of $j$ and $r/a$, which indicates a cross block degeneracy there. We show the contours on which Eq.~(\ref{crossblockcond}) is satisfied and therefore cross block degeneracies exist on our Segre plots in Fig. \ref{crossblockb}.

\begin{figure}[H]
    \centering
    \includegraphics{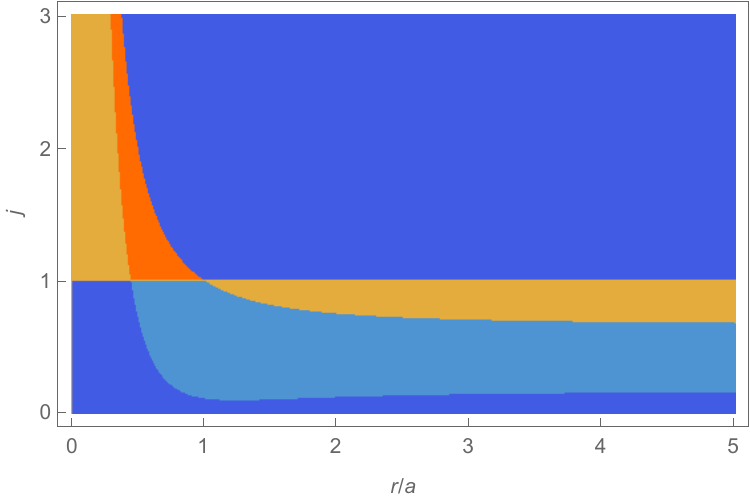}
    \includegraphics{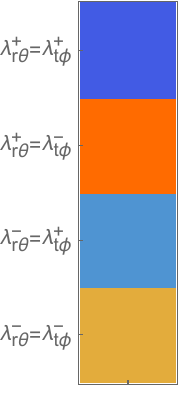}
    \caption{ This figure illustrates the particular cross-block degeneracies at $\chi^2=1$ for various $r/a,j$. We have $\lambda^-_{t\phi}=\lambda^-_{r\theta}$ in light orange (see legend to the right) and $\lambda^-_{t\phi}=\lambda^+_{r\theta}$ in dark orange, both of these indicate a [(1,1)] type degeneracy. We have $\lambda^+_{t\phi}=\lambda^+_{r\theta}$ in dark blue and $\lambda^+_{t\phi}=\lambda^-_{r\theta}$ in light blue, both of which are [(11)] cross block degeneracies. }
    \label{crossblocka}
\end{figure}

However, \eqref{crossblockcond} does not tell us whether it is a ``time space" or ``space space" pair between the blocks which is degenerate, i.e., whether the cross-block degeneracy involves a timelike vector of one block and a spacelike vector of the other, or both vectors are spacelike. In practice, for the example $m=0,j={\rm const}$ spacetime along $\chi^2=1$, the timelike eigenvector is $V_2$ in~\eqref{DSeigenvactors} and its corresponding eigenvalue is $\lambda_{t\phi}^-$. We can examine which pair of eigenvalues is degenerate for $\chi^2=1$. Eq.~\eqref{crossblockdef} with $\chi^2=1$ becomes 
\begin{align}
&
 \sign(j-1) \, z^2 \big[ (3 j+1) z^2+4\big]
+  \epsilon_{t\phi} \, (z^2+1) \, \Big| (3 j-2) z^2-1\Big| 
\nonumber\\&\qquad
- \epsilon_{r\theta} \, \Big|  (6 j-1) z^4+(3 j+1) z^2-1\Big| = 0.
\end{align}
When this is satisfied for $ \epsilon_{t\phi}=-1$ (timelike eigenvector) and either sign $ \epsilon_{r\theta}$ the cross block degeneracy is of the [(1,1)] type, when $ \epsilon_{t\phi}=1$ (spacelike eigenvector) it is of the [(11)] type. In Figure \ref{crossblocka}, we show which eigenvalues are degenerate along the $\chi^2=1$ for various $r$ and $j$.

 \subsection{Not all Segre types occur in this example}
 While this example with $m=0$ and positive constant $j\ne1$ shows a wide variety of Segre types, it does not show all the Segre types which could arise from combinations in Table \ref{segretable}. 
 For instance, the Segre types [2(11)] and its degenerate case [(211)] do not occur. 

 In order for the Segre type of the $t\phi$ block to be [2], we must not be at $\chi^2=1$ because then $\siiio=0$ and the Segre type of the $t\phi$ block will be [(1,1)]. The only points where the $r\theta$ block is [(11)] for which $\chi^2\ne1$ are those from \eqref{(11)a}. If one uses these values for $r$ and $\chi$, one obtains
 \begin{align}
    \Diiio\rightarrow \frac{\left(\sqrt{j}-1\right)^6 \left(3 j+44 \sqrt{j}+3\right)}{128 \pi ^2
   a^4 \left(\sqrt{j}+1\right)^4 j},
 \end{align}
 which can never be zero when $0\le j \le1/9$ which is required for the solution to \eqref{(11)a} to be within the domain of the coordinates. Therefore there is no manifestation of Segre [2(11)], which also precludes [(211)].

 \section{Equations of state}
 
The Kerr-Schild systems examined in Part I~\cite{Beltracchi2021a} are Segre type [(11)(1,1)] and hence automatically obey two equations of state between the four eigenvalues of the energy-momentum tensor. A further equation of state independent of position and time may then be cast into the form of a single equation between $\rho$ and $p_\perp$,
\begin{align}
F(\rho,p_\perp) = 0 .
\end{align}
Then the existence of such a relation is connected with the vanishing of the Jacobian determinant of the derivatives of $\rho$ and $p_\perp$ with respect to $r$ and $\theta$,
 \begin{equation}
 \label{eq:eos_condition}
     \frac{\partial p_\perp}{\partial r}\frac{\partial \rho}{\partial \theta}-\frac{\partial p_\perp}{\partial \theta}\frac{\partial \rho}{\partial r}=0
 \end{equation}
 for all $r$ and $\theta$. In Part I, this established a particular mass function $m(r)$ and a unique family of rotating Kerr-Schild systems that have the same equation of state in the rotating and nonrotating configurations. 
 
In the more general case discussed here, the situation is more complicated. 
Effectively, up to four equations of state 
\begin{align}
F_A(\lambda_1,\lambda_2,\lambda_3,\lambda_4)=0, \qquad A=1,\ldots,N,
\label{eq:FA}
\end{align}
with $N=1,2,3,$ or 4, may be found when there are relationships between the eigenvalues $\lambda_b$ ($b=1,\ldots,4$) that are independent of the position variables $r,\theta$ (independence of $t,\phi$ is guaranteed by our symmetries), and the same equations of state will apply regardless of rotation if the relationships between the eigenvalues is also independent of $a$. The $N$ equations of state~\eqref{eq:FA} can be thought geometrically as relations defining submanifolds in the four-dimensional space of eigenvalues $\lambda_b$. One equation $(N=1)$ defines a set of three-dimensional hypersurfaces (or volumes), two equations $(N=2)$ define a set of two-dimensional surfaces, three equations $(N=3)$ a set of lines, and four equations $(N=4)$ a discrete set of points. Each subspace can be parametrized in the form $\lambda_b(x_i)$ where $x_i=(r,\theta,a)$.  The dimension of the tangent space of the submanifolds equals the number of independent tangent vectors at a submanifold point, and this number equals the rank of the $4\times3$ matrix ${\partial \lambda_b}/{\partial x_i}$. In practice, it is easier to use the energy tensor invariants $I_b=(\Biiio,\BIii,\Diiio,\DIii)$, the submanifold coordinates $y_i=(r,\chi^2,a^2)$, and the rank of the $4\times3$ matrix
\begin{align}
    A_{bi}=\frac{\partial I_b}{\partial y_i} .
\end{align}

If all the components of $A_{bi}$ vanish, then the invariants $I_b$ are independent of the parameters $y_i$, the matrix $A_{bi}$ has rank 0, $I_b(y_i)$ describes a zero dimensional set, and there are 4 equations of state. We find this occurs for Drake-Szekeres systems if and only if
\begin{equation}
    j(r)=1,\qquad m(r)=M,
\end{equation}
where $M$ is a constant,
which is the standard Kerr solution. In this case, $I_b=(0,0,0,0)$, and the equations of state are $\lambda_1=\lambda_2=\lambda_3=\lambda_4=0$, which are the eigenvalues of the vacuum stress-energy tensor.

If all $2\times2$ minors of $A_{bi}$ vanish, but some of its components do not, then $A_{bi}$ has rank 1, $I_b(y_i)$ describes a one dimensional set, and there are three equations of state. We find this occurs for Drake-Szekeres systems if and only if
\begin{align}
    j(r)=1,\qquad m(r)=M-\frac{\Lambda c^2}{2 r}+\Lambda c r+\frac{\Lambda r^3}{6},
\end{align}
where $\Lambda,c,M$ are constants, which is the special Kerr-Schild system we describe in Paper 1 \cite{Beltracchi2021a}. In this case, $I_b(y_i)$ describes the line
\begin{align}
    \Biiio=\BIii,\qquad \Diiio=\DIii,\qquad \DIii=\BIii^2\left(1+\frac{4\pi \BIii}{\Lambda}\right)^2.
\end{align}
The equations of state are
\begin{align}
\lambda_{r\theta}^{+} = \lambda_{t\phi}^{+}, 
\quad
\lambda_{r\theta}^{-} = \lambda_{t\phi}^{-}, 
\quad
(\lambda_{r\theta}^{+} + \lambda_{r\theta}^{-} )^2 = -\frac{\Lambda}{2\pi} \lambda_{r\theta}^{\pm} ,
\end{align}
where in the last equation $\lambda_{r\theta}^{\pm} $ is $\lambda_{r\theta}^{+} $ or $ \lambda_{r\theta}^{-} $. This system is Segre type [(11)(1,1)] with an extra equation of state between the eigenvalue pairs.

If all $3\times3$ minors vanish but some $2\times2$ minors do not, then $A_{bi}$ has rank 2, $I_b(y_i)$ describes a two-dimensional set, and there are two equations of state. We have determined that this occurs when 
\begin{align}
    j(r)=1
\end{align}
for arbitrary $m(r)$, in which case $I_b(y_i)$ describes the plane
\begin{align}
    \Biiio=\BIii,\qquad \Diiio=\DIii,
\end{align}
and the equations of state are
\begin{align}
\lambda_{r\theta}^{\pm} = \lambda_{t\phi}^{\pm}, 
\end{align}
so it has Segre type [(11)(1,1)]. The case just described is only a sufficient condition for the matrix $A_{bi}$ to have rank 2. We have determined that the most general function $j(r)$ for which all $3\times3$ minors of $A_{bi}$ vanish must satisfy a complicated nonlinear differential equation of which we have not found the solution. 

If none of the $3\times3$ minors of $A_{bi}$ vanish, then $A_{bi}$ has rank $3$,
$I_b(y_i)$ describes a 3-dimensional space, and there is 1 equation of state, although finding it in a generic case seems to be extremely complicated. For example, in case $j(r)$ and $m(r)$ are given functions, one can in principle eliminate $y_i$ from the equations $I_b=I_b(y_i)$. However, even in the simple case $m=0, j={\rm const}$ of the ``rotating Minkowski'' spacetime, the elimination procedure is very complicated and leads to cumbersome expressions. 

 \section{Conclusions}

 The Newman-Janis algorithm can be used to create the Kerr and Kerr-Newman metrics from the Schwarzschild and Reissner-Nordstrom metrics. Additionally, its generalizations allow for the construction of rotating systems which reduce to spherical systems in the limit of no rotation. These distortions of Segre type and the equations of state would require explanations from a fundamental theory of the matter content in the systems. One should therefore be careful in identifying a Newman-Janis system with a physically rotating version of the seed system.

 In the Gurses-Gursey generalization for Kerr-Schild systems has been discussed in Part I~\cite{Beltracchi2021a}, where a unique family of stationary axisymmetric Kerr-Schild systems was identified as having the same equation of state both for the rotating and nonrotating configurations. 
 
 The Drake-Szekeres generalization allows for usage of non Kerr-Schild metrics. The energy-momentum tensors are significantly more complicated in general than the Gurses-Gursey type and they can be any Segre type but $[31]$ and $[(31)]$. They can for example feature severe distortions of Segre type such as generation of systems containing $[11Z\bar{Z}]$ from seed systems which were initially [(111,1)], which we explicitly show in our analysis of the ``spinning Minkowski" space. 
 We have analyzed the existence of equations of state independent of position and rotation for Drake-Szekeres systems. We find that a Drake-Szekeres system with 4 equations of state is the Kerr spacetime, that with 3 equations of state it must be the special Kerr-Schild system we found in Part I, and that with two equations of state it is either a general Kerr-Schild system or possibly something with $j(r)$, $m(r)$ which satisfy complicated nonlinear differential equations. Finally, by counting arguments of having four eigenvalues with each a function of three parameters, there should be one equation of state in general.

\acknowledgments
 P.G. was partially supported by NSF grant PHY-2014075, and is very grateful to Prof.\ Masahide Yamaguchi for his generous support under JSPS Grant-in-Aid for Scientific Research Number JP18K18764 at the Tokyo Institute of Technology.
\appendix
\renewcommand\theequation{\Alph{section}.\arabic{equation}}

\section{Components of the energy-momentum tensor}

In this appendix we list the explicit expressions of the quantities $\hat{\mu}_i$ and $\hat{\mu}_{ij}$ appearing in the components of the energy-momentum tensor and in its eigenvalues. Notice that they are rational functions of $\chi=\cos\theta$, and are structured as a prefactor times a polynomial in $\chi$. They are written here as polynomial functions in $m$, $m'$, $m''$, $j'$, and $j''$.
\begin{subequations}
\label{eq:mus}
\begin{align}
    & \muo=\frac{1}{32\pi \Sigma^3\Sigma_j^2}\Bigbra
    -8r^2\Sigma\Sigma_j^2 m'
    + (1-\chi^2) a^2 r^3 \Sigma^2 (4j+rj') j' 
    \nonumber\\&     \qquad\ \
    + 8(j-1)a^2r^2\big[(1-\chi^2)(r^4-a^4\chi^4)-3r\chi^2(\Sigma+\Sigma_j) m\big]
    \nonumber\\&     \qquad\ \
    + 4(j-1)^2 a^2r^2\Sigma\big[(1-\chi^2)\Sigma+r^2(1-4\chi^2)\big]
    \Bigket
\\
    & \mui=\frac{1}{32\pi \Sigma^3\Sigma_j^2}\Bigbra-8r^2\Sigma \Sigma_j^2m' +a^2r^4(1-\chi^2)\Sigma^2(j')^2\nonumber\\&-4r^3\Sigma j'\Big[2a^2\chi^2(a^2+r^2)+\big(\Sigma^2+(a^2+r^2)(r^2-a^2\chi^2)\big)j-4 m r \Sigma_j\Big]+\nonumber\\&
     8 a^2 r^2 \Sigma (j-1)\Big[(\chi^2-5)\Sigma+6r^2(1-\chi^2)+2r\chi^2 m\Big]-\nonumber\\& 4a^2r^2(j-1)^2\Big[6r^4(\chi^2-1)+r^2\Sigma(5+4\chi^2)+\Sigma^2(\chi^2-1)-6r^3\chi^2 m\Big]
     \Bigket
\\
       &\muii=\frac{1}{32\pi \Sigma^3\Sigma_j^2}\Bigbra
  -4 r \Sigma^2 \Sigma_j^2 m'' -8a^2\chi^2\Sigma^3 m'
  - 4 r^2 \Sigma^2 \Sigma_j (r^2+a^2-2rm) j''
  +
  \nonumber\\&
  r^4 \Sigma^2 \big[ 7(a^2+r^2) + \Sigma-16 r m \big] (j')^2
  + 4 r \Sigma j' \big\{ r m \left[-j r^2 \left(5 a^2 \chi ^2+r^2\right)+15 a^2 r^2 \chi ^2+11 a^4 \chi
   ^4\right]
   \nonumber\\&
   +j r^2 \left[a^2 r^2 \left(2 \chi ^2+1\right)+a^4 \chi ^2 \left(\chi
   ^2+3\right)-r^4\right]-a^2 \chi ^2 \left[a^2 r^2 \left(7 \chi ^2+6\right)+4 a^4 \chi
   ^2+9 r^4\right]+3 r^2 \Sigma  \Sigma_j m' \big\}
    \nonumber\\&    
  +8(j-1)a^2\chi^2  \big\{\Sigma  \left[m' r^2 \left(a^2 \chi ^2+2 j r^2-r^2\right)-5 a^2 r^2 \left(\chi
   ^2-1\right)-a^4 \chi ^2+r^4\right]+r m \big[3 a^2 r^2 \chi ^2+5 a^4 \chi ^4\nonumber\\&+r^4-3 j (a^2 r^2 \chi ^2+2
   r^4)\big]\big\}
  + 4 (j-1)^2 (3r^4+11a^2r^2+5a^4\chi^2-2a^2r^2\chi^2+a^4\chi^4) a^2r^2\chi^2
    \Bigket
\\
  & \muiii=\frac{1}{32\pi \Sigma^3\Sigma_j^2}\Bigbra
  -4r\Sigma^2\Sigma_j^2m'' 
  -8 a^2 \chi^2 \Sigma \Sigma_j (3\Sigma-2\Sigma_j) m'
  +12r^3\Sigma^2\Sigma_j j'm'
  \nonumber\\&
  - 16 r^5 \Sigma^2 m (j')^2
  + 8 r^3 \Sigma^2 \Sigma_j m j''
  - 4r^2\Sigma [jr^4-a^4\chi^4+5a^2\chi^2(\Sigma_j-3\Sigma)] m j'
  \nonumber\\&
  + 8 a^2 r \chi^2 (j-1) [r^4-a^4\chi^4-3a^2r^2\chi^2(j-1)-6(jr^4-a^4\chi^4)]m
  - 4r^2 (a^2+r^2) \Sigma^2 \Sigma_j j''
  \nonumber\\&
  + r^4 \Sigma^2 [5a^2(1-\chi^2)+8\Sigma] (j')^2
  \nonumber\\&
  - 4 r \Sigma (6a^4r^2\chi^2+9a^2r^4\chi^2+4a^6\chi^4+7a^4r^2\chi^4+a^2r^4j+r^6j-a^4jr^2\chi^2-4a^2r^4\chi^2j-3a^4r^2\chi^4j) j'
  \nonumber\\&
  -4a^2\Sigma \chi^2(j-1)(a^2r^2+r^4+2a^4\chi^2+5a^2r^2\chi^2-3a^2jr^2-3r^4j-3a^2r^2\chi^2j)
  \Bigket
\\
    & \muiiio=\frac{a}{16 \pi \Sigma ^2 \Sigma_j^2 }\Bigbra
    -2r(r^2 \Sigma_j+2\Sigma a^2\chi^2 )  j'
    + \Sigma r^4 (j')^2 
    - \Sigma \Sigma_j r^2 j''
    - 2 (j-1) (r^2-a^2\chi^2)(jr^2-a^2\chi^2)
    \Bigket
    \label{sigma30}
\\
   &\muIii =\frac{3a^2r \cos\theta }{8\pi \Sigma ^3\Sigma_j^2} \Bigbra
   2(j-1)(a^4 \chi^4-jr^4) +\Sigma^2 r j' 
   \Bigket
   \label{sigmainj}
\end{align}
\end{subequations}

We can see that both a nonconstant $j$ and a $j\ne1$ contribute many terms. When $j=1$, the quantities $\muo,\mui,\muii,\muiii$ go to the forms in Eqs.~(\ref{eq:spinDENP}), and $\muIii=\muiiio=0$.

The coordinate components of the energy-momentum tensor have expressions independent of the sign of $\Delta$ (they are polynomial functions of $\Delta$ and its derivatives $\Delta'$, $\Delta''$),
\begin{subequations}
\begin{align}
    &T_{tt}=\frac{\Sigma}{\Sigma_j^2}   \big( a^2 \muiii \sin^2\theta- 2a \sin^2\theta \muiiio \Delta-\muo  \Delta    \big) \\
    &T_{t\phi}=\frac{\Sigma\sin^2\theta}{\Sigma_j^2}   \big[   (a^2+r^2 j+a^2\sin^2\theta)\muiiio \Delta-a \muiii
   (a^2+r^2 j)+a \muo  \Delta   \big],\\
   &T_{\phi\phi}=-\frac{\Sigma\sin^2\theta}{\Sigma_j^2} \big[a^2  \muo  \Delta \sin^2\theta +2 a \sin^2\theta
   (a^2+r^2 j)\muiiio \Delta-\muiii (a^2+r^2 j)^2\big],\\
   &T_{rr}=\frac{\Sigma \mui}{\Delta },\\
   &T_{r\theta}=  \Sigma \muIii \sin\theta,\\
   &T_{\theta\theta}= \Sigma \muii.
\end{align}
\end{subequations}
For completeness, we also list the expressions of the mixed components $T^\mu_{~\nu}$, which appear for example in the invariant traces and discriminants $\BIii$, $\Biiio$, $\DIii$, and $\Diiio$ in Eqs.~\eqref{eq:BD_covariant}.
\begin{subequations}
\begin{align}
    &T^t_{~t}=-\frac{1}{\Sigma_j}   \big[ a^2 \muiii \sin^2\theta- a \sin^2\theta (\Delta+a^2+jr^2) \muiiio -\muo  (a^2+r^2 j)  \big]\\
    &T^t_{~\phi}= \frac{\sin^2\theta}{\Sigma_j} \left[  a(a^2+jr^2)(\muiii-\muo)-\big[a^2 \Delta \sin^2\theta+(a^2+jr^2)^2 \big] \muiiio   \right] \\
    &T^\phi_{~t}= \frac{1}{\Sigma_j} \left[ (\Delta+a^2\sin^2\theta ) \muiiio -a (\muiii -\muo)  \right]  \\
   &T^\phi_{~\phi}=\frac{1}{\Sigma_j} \big[-a^2  \muo  \sin^2\theta - a \sin^2\theta (\Delta+a^2+jr^2)  \muiiio + \muiii (a^2+r^2 j)\big],\\
   &T^r_{~r}=\mui \\
   &T^r_{~\theta}=  \muIii  \Delta \sin\theta,\\
   &T^\theta_{~r}=   \muIii \sin\theta , \\
   &T^\theta_{~\theta}=  \muii.
\end{align}
\end{subequations}
\section{Technical details of equation of state calculation}

To find the conditions on the existence and number of equations of state that DS systems can have, we have used a \textit{Mathematica} code that we explain in this appendix. This code uses some special procedures to obtain the result, as the usage of standard manipulation functions like {\tt Expand}, {\tt Collect}, and {\tt Together} fail to produce a result in a reasonable amount of time on the complicated rational expressions appearing in the calculation. We use version 12.2.0 of \textit{Mathematica}~\cite{Mathematica} on a 2020 Apple M1 MacBook Pro with 16 GB of RAM. All lengthy output was suppressed to shorten the execution time.

The input quantities {\tt B12e, B30e, D12e, D30e}, which have been precomputed and whose expressions we do not include here, are equal
\begin{align}
{\tt B12e} = p \, \BIii,
\qquad 
{\tt B30e} = p \, \Biiio 
\qquad
{\tt D12e} = p^2 \, \DIii, 
\qquad 
{\tt D30e} = p^2 \, \Diiio, 
\end{align}
where the invariants $\BIii,\Biiio,\DIii,\Diiio$ follow from Eqs.~\eqref{eq:BB} and~\eqref{eq:DD} after inserting the expressions of $\muo,\mui,\muii,\muiii,\muIii,\muiiio$ in~\eqref{eq:mus}, and the factor
\begin{align}
p=32\pi \Sigma^3\Sigma_j^2
\end{align}
is introduced to make {\tt B12e, B30e, D12e, D30e} polynomials in $a^2$, $\chi^2$, $j(r)$, $m(r)$ and the derivatives of the latter two functions. To exploit this polynomial dependence, the code uses the variables 
\begin{align}
& {\tt D0j}=j,  
&& {\tt D1j}=j',  
&& {\tt D2j}=j'',  
&& {\tt D3j}=j'''; 
\\
& {\tt D0m}=m,  
&& {\tt D1m}=m',  
&& {\tt D2m}=m'',  
&& {\tt D3m}=m'''.
\end{align}
The matrix of tangent vectors $A_{bi} = \partial I_b/\partial y_i$ of the three-dimensional surface in the four-dimensional space of the invariants $I_b = (\BIii,\Biiio,\DIii,\Diiio)$, with $y_i=(r,\chi^2,a^2)$, is represented in the code by a matrix
\begin{align}
{\tt tanvectem} = p \, \Sigma\, \Sigma_j \, A_{bi} , 
\end{align}
To avoid nonpolynomial manipulations, the $\BIii$ and $\Biiio$ components of these vectors are computed as 
\begin{align}
{\tt Factor[ipref \ D[pref,}y_i{\tt ]] \ B12e + ipref \ pref \ D[B12e,}y_i{\tt ] }
\end{align} 
and their $\DIii$ and $\Diiio$ components as 
\begin{align}
{\tt Factor[ipref^2 \ D[pref^2,}y_i{\tt ]] \ D12e + ipref^2 \ pref^2 \ D[D12e,}y_i{\tt ] },
\end{align} 
where ${\tt pref}=1/(32\pi\Sigma^3\Sigma_j^2)$ and ${\tt ipref}=32\pi\Sigma^4\Sigma_j^3$.

At this stage, we have polynomial expressions ${\tt tanvectem}$ for the tangent vectors in the variables ${\tt aa}=a^2$, $\chi\chi=\chi^2$, ${\tt D0j}$, ${\tt D1j}$, ${\tt D2j}$, ${\tt D3j}$, ${\tt D0m}$, ${\tt D1m}$, ${\tt D2m}$, ${\tt D3m}$. The computational problem is now to find the minors of the $4\times 3$ matrix ${\tt tanvectem}$,
extract its coefficients in ${\tt aa}$ and $\chi\chi$, and find the algebraic conditions on the variables ${\tt D0j}$, ${\tt D1j}$, ${\tt D2j}$, ${\tt D3j}$, ${\tt D0m}$, ${\tt D1m}$, ${\tt D2m}$, ${\tt D3m}$, i.e., the differential equations for $j(r)$ and $m(r)$, that make all those coefficients vanish simultaneously. 

Since we are looking for products of the matrix elements to be zero simultaneously, we pull out factors in each matrix element that cannot vanish for all values of $y_i$, giving us two matrices ${\tt vsf}$ and ${\tt vlf}$ containing the pulled-out factors and the remaining factors, respectively. Then we extract the coefficients of ${\tt aa}$ and $\chi\chi$ in  ${\tt vsf}$ and ${\tt vlf}$ using
\begin{align}
{\tt vlf = Map[CoefficientList[\#, \{aa, \chi\chi\}, \{13, 13\}] \&,{\tt vem}, \{2\}] } ,
\end{align}
and similarly for ${\tt vsf}$, where the value 13 is large enough to include all powers of the variables. We do this in order to compute the coefficients of ${\tt aa}$ and $\chi\chi$ in the minors by means of ${\tt ListConvolve}$ rather than direct polynomial multiplication, which was too slow. For this purpose we defined a function
\begin{align}
{\tt listtimes2[p1\_,p2\_] := ListConvolve[p1,p2,\{1,-1\},0]}
\end{align}
for the product of two polynomials represented by their coefficients, and analogous functions for the product of three polynomials and the determinants of matrices with polynomial elements.

We then proceed to analyze the cases of $1\times1$, $2\times2$, and $3\times3$ minors one at a time. 
The $1\times1$ minors are simply the elements of the matrix ${\tt vlf}$, since the factors in ${\tt vsf}$ do not vanish identically,
\begin{align}
{\tt minors1 = vlf} .
\end{align}
We select the terms in ${\tt minors1}$ that do not contain the mass function $m(r)$ or its derivatives, and thus are functions of $r$, $j(r)$ and the derivatives of $j(r)$ only. To do this, we avoid the use of the Mathematica function ${\tt FreeQ}$, which was too slow, and use instead a sequence of code lines of the form
\begin{align}
& {\tt tmp=Select[tmp,Coefficient[\#,}DXm{\tt ,}e{\tt ]==0 \&]]}
\label{eq:B11}
\end{align}
where the exponent $e$ ranges successively from 6 to 1, and $DXm$ is ${\tt D3m, D2m, \ldots, D0m}$.
Setting the result of this to zero, using the function ${\tt Reduce}$ on the resulting equalities, followed by ${\tt FullSimplify}$, produces the result
\begin{align}
{\tt D0j == 1 \ \&\& \ D1j == 0 \ \&\& \ D2j == 0}
\end{align}
which means that the only case with a third equation of state is $j(r)=1$. We then introduce this function $j(r)$ back into all the minors ${\tt minors1}$,  and solve the resulting equations ${\tt minors1}=0$ for $DXm$ by means of the ${\tt Reduce}$ function. This gives 
\begin{align}
{\tt D0m == 1 \ \&\& \ D1m == 0 \ \&\& \ D2m == 0 \ \&\& \ D3m == 0},
\end{align}
that is the only solution is $m(r)=M={\rm const}$. Viceversa, it is easy to verify that $j(r)=1$ and $m(r)=M={\rm const}$ imply that all $1\times1$ minors vanish. Therefore all $1\times1$ minors of $A_{bi}$ vanish if and only if $j(r)=1$ and $m(r)=M={\rm const}$.

For the $2\times2$ minors, we first pull out of each minor the factors that do not vanish identically by formally computing the $2\times2$ minors of a matrix ${\tt vsf} * {\tt elf}$, where ${\tt elf}$ is a placeholder matrix, dropping nonidentically vanishing factors in each minor, and then replacing ${\tt elf}$ with the actual matrix ${\tt vlf}$. The distinct elements of the resulting array of minors is collected in the \textit{Mathematica} list ${\tt minors2}$. We extract from it the terms that do not contain $m(r)$ and its derivatives, by the same procedure described in Eq.~\eqref{eq:B11}. Applying ${\tt Reduce}$ to the remaining terms does not produce a result in a reasonable amount of time. 
We split each element in ${\tt minors2}$ without $m(r)$ and its derivatives into its factors, and impose that one of the factors is zero. This gives us conditions on $DXj$, which we find have the solution $j(r)=1$ only. Using the latter in the full ${\tt minors2}$ matrix gives the differential equation
\begin{align}
   4m'^2-2 r m' m''+ r^2 m''^2 -2r^2 m' m'''=0,
\end{align}
which is the same as Eq. (5.6) in part I. This shows that the $2\times2$ minors only vanish for the special family of Kerr-Schild systems we examine in part 1.

For the $3\times3$ minors, we proceed in a similar way to the $2\times2$ minors by usage of the ${\tt elf}$ placeholder matrix, dropping nonvanishing factors, and looking for terms which do not contain $DXm$ to find conditions on $j$. The conditions on $j$ we find are: $j=1$, which works for any $m$; $j=1+c/r^2$, for which no $m$ will work when $c\ne0$; and complicated differential equations for $j$ for which we have not found the explicit form of $j$ or $m$.
\bibliographystyle{apsrev4-2}
\bibliography{monster.bib}

\begin{thebibliography}{30}%
\makeatletter
\providecommand \@ifxundefined [1]{%
 \@ifx{#1\undefined}
}%
\providecommand \@ifnum [1]{%
 \ifnum #1\expandafter \@firstoftwo
 \else \expandafter \@secondoftwo
 \fi
}%
\providecommand \@ifx [1]{%
 \ifx #1\expandafter \@firstoftwo
 \else \expandafter \@secondoftwo
 \fi
}%
\providecommand \natexlab [1]{#1}%
\providecommand \enquote  [1]{``#1''}%
\providecommand \bibnamefont  [1]{#1}%
\providecommand \bibfnamefont [1]{#1}%
\providecommand \citenamefont [1]{#1}%
\providecommand \href@noop [0]{\@secondoftwo}%
\providecommand \href [0]{\begingroup \@sanitize@url \@href}%
\providecommand \@href[1]{\@@startlink{#1}\@@href}%
\providecommand \@@href[1]{\endgroup#1\@@endlink}%
\providecommand \@sanitize@url [0]{\catcode `\\12\catcode `\$12\catcode
  `\&12\catcode `\#12\catcode `\^12\catcode `\_12\catcode `\%12\relax}%
\providecommand \@@startlink[1]{}%
\providecommand \@@endlink[0]{}%
\providecommand \url  [0]{\begingroup\@sanitize@url \@url }%
\providecommand \@url [1]{\endgroup\@href {#1}{\urlprefix }}%
\providecommand \urlprefix  [0]{URL }%
\providecommand \Eprint [0]{\href }%
\providecommand \doibase [0]{https://doi.org/}%
\providecommand \selectlanguage [0]{\@gobble}%
\providecommand \bibinfo  [0]{\@secondoftwo}%
\providecommand \bibfield  [0]{\@secondoftwo}%
\providecommand \translation [1]{[#1]}%
\providecommand \BibitemOpen [0]{}%
\providecommand \bibitemStop [0]{}%
\providecommand \bibitemNoStop [0]{.\EOS\space}%
\providecommand \EOS [0]{\spacefactor3000\relax}%
\providecommand \BibitemShut  [1]{\csname bibitem#1\endcsname}%
\let\auto@bib@innerbib\@empty
%</preamble>
\bibitem [{\citenamefont {Newman}\ and\ \citenamefont
  {Janis}(1965)}]{Newman:1965tw}%
  \BibitemOpen
  \bibfield  {author} {\bibinfo {author} {\bibfnamefont {E.~T.}\ \bibnamefont
  {Newman}}\ and\ \bibinfo {author} {\bibfnamefont {A.~I.}\ \bibnamefont
  {Janis}},\ }\href {https://doi.org/10.1063/1.1704350} {\bibfield  {journal}
  {\bibinfo  {journal} {J. Math. Phys.}\ }\textbf {\bibinfo {volume} {6}},\
  \bibinfo {pages} {915} (\bibinfo {year} {1965})}\BibitemShut {NoStop}%
%%CITATION = JMAPA,6,915;%%
\bibitem [{\citenamefont {Newman}\ \emph {et~al.}(1965)\citenamefont {Newman},
  \citenamefont {Couch}, \citenamefont {Chinnapared}, \citenamefont {Exton},
  \citenamefont {Prakash},\ and\ \citenamefont {Torrence}}]{Newman:1965my}%
  \BibitemOpen
  \bibfield  {author} {\bibinfo {author} {\bibfnamefont {E.~T.}\ \bibnamefont
  {Newman}}, \bibinfo {author} {\bibfnamefont {R.}~\bibnamefont {Couch}},
  \bibinfo {author} {\bibfnamefont {K.}~\bibnamefont {Chinnapared}}, \bibinfo
  {author} {\bibfnamefont {A.}~\bibnamefont {Exton}}, \bibinfo {author}
  {\bibfnamefont {A.}~\bibnamefont {Prakash}},\ and\ \bibinfo {author}
  {\bibfnamefont {R.}~\bibnamefont {Torrence}},\ }\href
  {https://doi.org/10.1063/1.1704351} {\bibfield  {journal} {\bibinfo
  {journal} {J. Math. Phys.}\ }\textbf {\bibinfo {volume} {6}},\ \bibinfo
  {pages} {918} (\bibinfo {year} {1965})}\BibitemShut {NoStop}%
%%CITATION = JMAPA,6,918;%%
\bibitem [{\citenamefont {{Kerr}}(2007)}]{Kerr:2007dk}%
  \BibitemOpen
  \bibfield  {author} {\bibinfo {author} {\bibfnamefont {R.~P.}\ \bibnamefont
  {{Kerr}}}\ }(\bibinfo {year} {2007})\ p.\ \bibinfo {pages}
  {arXiv:0706.1109},\ \Eprint {https://arxiv.org/abs/0706.1109}
  {arXiv:0706.1109 [gr-qc]} \BibitemShut {NoStop}%
\bibitem [{\citenamefont {Gurses}\ and\ \citenamefont
  {Gursey}(1975)}]{Gurses:1975vu}%
  \BibitemOpen
  \bibfield  {author} {\bibinfo {author} {\bibfnamefont {M.}~\bibnamefont
  {Gurses}}\ and\ \bibinfo {author} {\bibfnamefont {F.}~\bibnamefont
  {Gursey}},\ }\href {https://doi.org/10.1063/1.522480} {\bibfield  {journal}
  {\bibinfo  {journal} {J. Math. Phys.}\ }\textbf {\bibinfo {volume} {16}},\
  \bibinfo {pages} {2385} (\bibinfo {year} {1975})}\BibitemShut {NoStop}%
%%CITATION = JMAPA,16,2385;%%
\bibitem [{\citenamefont {Smailagic}\ and\ \citenamefont
  {Spallucci}(2010)}]{Smailagic:2010nv}%
  \BibitemOpen
  \bibfield  {author} {\bibinfo {author} {\bibfnamefont {A.}~\bibnamefont
  {Smailagic}}\ and\ \bibinfo {author} {\bibfnamefont {E.}~\bibnamefont
  {Spallucci}},\ }\href {https://doi.org/10.1016/j.physletb.2010.03.075}
  {\bibfield  {journal} {\bibinfo  {journal} {Phys. Lett.}\ }\textbf {\bibinfo
  {volume} {B688}},\ \bibinfo {pages} {82} (\bibinfo {year} {2010})},\ \Eprint
  {https://arxiv.org/abs/1003.3918} {arXiv:1003.3918 [hep-th]} \BibitemShut
  {NoStop}%
%%CITATION = ARXIV:1003.3918;%%
\bibitem [{\citenamefont {Bambi}\ and\ \citenamefont
  {Modesto}(2013)}]{Bambi:2013ufa}%
  \BibitemOpen
  \bibfield  {author} {\bibinfo {author} {\bibfnamefont {C.}~\bibnamefont
  {Bambi}}\ and\ \bibinfo {author} {\bibfnamefont {L.}~\bibnamefont
  {Modesto}},\ }\href {https://doi.org/10.1016/j.physletb.2013.03.025}
  {\bibfield  {journal} {\bibinfo  {journal} {Phys. Lett.}\ }\textbf {\bibinfo
  {volume} {B721}},\ \bibinfo {pages} {329} (\bibinfo {year} {2013})},\ \Eprint
  {https://arxiv.org/abs/1302.6075} {arXiv:1302.6075 [gr-qc]} \BibitemShut
  {NoStop}%
%%CITATION = ARXIV:1302.6075;%%
\bibitem [{\citenamefont {Ghosh}(2015)}]{Ghosh:2014pba}%
  \BibitemOpen
  \bibfield  {author} {\bibinfo {author} {\bibfnamefont {S.~G.}\ \bibnamefont
  {Ghosh}},\ }\href {https://doi.org/10.1140/epjc/s10052-015-3740-y} {\bibfield
   {journal} {\bibinfo  {journal} {Eur. Phys. J.}\ }\textbf {\bibinfo {volume}
  {C75}},\ \bibinfo {pages} {532} (\bibinfo {year} {2015})},\ \Eprint
  {https://arxiv.org/abs/1408.5668} {arXiv:1408.5668 [gr-qc]} \BibitemShut
  {NoStop}%
%%CITATION = ARXIV:1408.5668;%%
\bibitem [{\citenamefont {Dymnikova}\ and\ \citenamefont
  {Galaktionov}(2015)}]{Dymnikova:2015hka}%
  \BibitemOpen
  \bibfield  {author} {\bibinfo {author} {\bibfnamefont {I.}~\bibnamefont
  {Dymnikova}}\ and\ \bibinfo {author} {\bibfnamefont {E.}~\bibnamefont
  {Galaktionov}},\ }\href {https://doi.org/10.1088/0264-9381/32/16/165015}
  {\bibfield  {journal} {\bibinfo  {journal} {Class. Quant. Grav.}\ }\textbf
  {\bibinfo {volume} {32}},\ \bibinfo {pages} {165015} (\bibinfo {year}
  {2015})},\ \Eprint {https://arxiv.org/abs/1510.01353} {arXiv:1510.01353
  [gr-qc]} \BibitemShut {NoStop}%
%%CITATION = ARXIV:1510.01353;%%
\bibitem [{\citenamefont {Ghosh}(2016)}]{Ghosh:2015ovj}%
  \BibitemOpen
  \bibfield  {author} {\bibinfo {author} {\bibfnamefont {S.~G.}\ \bibnamefont
  {Ghosh}},\ }\href {https://doi.org/10.1140/epjc/s10052-016-4051-7} {\bibfield
   {journal} {\bibinfo  {journal} {Eur. Phys. J.}\ }\textbf {\bibinfo {volume}
  {C76}},\ \bibinfo {pages} {222} (\bibinfo {year} {2016})},\ \Eprint
  {https://arxiv.org/abs/1512.05476} {arXiv:1512.05476 [gr-qc]} \BibitemShut
  {NoStop}%
%%CITATION = ARXIV:1512.05476;%%
\bibitem [{\citenamefont {Atamurotov}\ \emph {et~al.}(2016)\citenamefont
  {Atamurotov}, \citenamefont {Ghosh},\ and\ \citenamefont
  {Ahmedov}}]{Atamurotov:2015xfa}%
  \BibitemOpen
  \bibfield  {author} {\bibinfo {author} {\bibfnamefont {F.}~\bibnamefont
  {Atamurotov}}, \bibinfo {author} {\bibfnamefont {S.~G.}\ \bibnamefont
  {Ghosh}},\ and\ \bibinfo {author} {\bibfnamefont {B.}~\bibnamefont
  {Ahmedov}},\ }\href {https://doi.org/10.1140/epjc/s10052-016-4122-9}
  {\bibfield  {journal} {\bibinfo  {journal} {Eur. Phys. J.}\ }\textbf
  {\bibinfo {volume} {C76}},\ \bibinfo {pages} {273} (\bibinfo {year}
  {2016})},\ \Eprint {https://arxiv.org/abs/1506.03690} {arXiv:1506.03690
  [gr-qc]} \BibitemShut {NoStop}%
%%CITATION = ARXIV:1506.03690;%%
\bibitem [{\citenamefont {Lamy}\ \emph {et~al.}(2018)\citenamefont {Lamy},
  \citenamefont {Gourgoulhon}, \citenamefont {Paumard},\ and\ \citenamefont
  {Vincent}}]{Lamy:2018zvj}%
  \BibitemOpen
  \bibfield  {author} {\bibinfo {author} {\bibfnamefont {F.}~\bibnamefont
  {Lamy}}, \bibinfo {author} {\bibfnamefont {E.}~\bibnamefont {Gourgoulhon}},
  \bibinfo {author} {\bibfnamefont {T.}~\bibnamefont {Paumard}},\ and\ \bibinfo
  {author} {\bibfnamefont {F.~H.}\ \bibnamefont {Vincent}},\ }\href
  {https://doi.org/10.1088/1361-6382/aabd97} {\bibfield  {journal} {\bibinfo
  {journal} {Class. Quant. Grav.}\ }\textbf {\bibinfo {volume} {35}},\ \bibinfo
  {pages} {115009} (\bibinfo {year} {2018})},\ \Eprint
  {https://arxiv.org/abs/1802.01635} {arXiv:1802.01635 [gr-qc]} \BibitemShut
  {NoStop}%
%%CITATION = ARXIV:1802.01635;%%
\bibitem [{\citenamefont {Sakti}\ \emph {et~al.}(2019)\citenamefont {Sakti},
  \citenamefont {Prihadi}, \citenamefont {Suroso},\ and\ \citenamefont
  {Zen}}]{Sakti:2019iku}%
  \BibitemOpen
  \bibfield  {author} {\bibinfo {author} {\bibfnamefont {M.~F. A.~R.}\
  \bibnamefont {Sakti}}, \bibinfo {author} {\bibfnamefont {H.~L.}\ \bibnamefont
  {Prihadi}}, \bibinfo {author} {\bibfnamefont {A.}~\bibnamefont {Suroso}},\
  and\ \bibinfo {author} {\bibfnamefont {F.~P.}\ \bibnamefont {Zen}}\
  }(\bibinfo {year} {2019})\ \Eprint {https://arxiv.org/abs/1911.07569}
  {arXiv:1911.07569 [gr-qc]} \BibitemShut {NoStop}%
%%CITATION = ARXIV:1911.07569;%%
\bibitem [{\citenamefont {Shaikh}(2019)}]{PhysRevD.100.024028}%
  \BibitemOpen
  \bibfield  {author} {\bibinfo {author} {\bibfnamefont {R.}~\bibnamefont
  {Shaikh}},\ }\href {https://doi.org/10.1103/PhysRevD.100.024028} {\bibfield
  {journal} {\bibinfo  {journal} {Phys. Rev. D}\ }\textbf {\bibinfo {volume}
  {100}},\ \bibinfo {pages} {024028} (\bibinfo {year} {2019})}\BibitemShut
  {NoStop}%
\bibitem [{\citenamefont {{Beltracchi}}\ and\ \citenamefont
  {{Gondolo}}(2021)}]{Beltracchi2021a}%
  \BibitemOpen
  \bibfield  {author} {\bibinfo {author} {\bibfnamefont {P.}~\bibnamefont
  {{Beltracchi}}}\ and\ \bibinfo {author} {\bibfnamefont {P.}~\bibnamefont
  {{Gondolo}}},\ }\href@noop {} {\bibfield  {journal} {\bibinfo  {journal}
  {arXiv e-prints}\ ,\ \bibinfo {eid} {arXiv:2104.02255}} (\bibinfo {year}
  {2021})},\ \Eprint {https://arxiv.org/abs/2104.02255} {arXiv:2104.02255
  [gr-qc]} \BibitemShut {NoStop}%
\bibitem [{\citenamefont {Drake}\ and\ \citenamefont
  {Szekeres}(2000)}]{Drake:1998gf}%
  \BibitemOpen
  \bibfield  {author} {\bibinfo {author} {\bibfnamefont {S.~P.}\ \bibnamefont
  {Drake}}\ and\ \bibinfo {author} {\bibfnamefont {P.}~\bibnamefont
  {Szekeres}},\ }\href {https://doi.org/10.1023/A:1001920232180} {\bibfield
  {journal} {\bibinfo  {journal} {Gen. Rel. Grav.}\ }\textbf {\bibinfo {volume}
  {32}},\ \bibinfo {pages} {445} (\bibinfo {year} {2000})},\ \Eprint
  {https://arxiv.org/abs/gr-qc/9807001} {arXiv:gr-qc/9807001 [gr-qc]}
  \BibitemShut {NoStop}%
%%CITATION = GR-QC/9807001;%%
\bibitem [{\citenamefont {Lombardo}(2004)}]{Lombardo_2004}%
  \BibitemOpen
  \bibfield  {author} {\bibinfo {author} {\bibfnamefont {D.~J.~C.}\
  \bibnamefont {Lombardo}},\ }\href
  {https://doi.org/10.1088/0264-9381/21/6/009} {\bibfield  {journal} {\bibinfo
  {journal} {Classical and Quantum Gravity}\ }\textbf {\bibinfo {volume}
  {21}},\ \bibinfo {pages} {1407} (\bibinfo {year} {2004})}\BibitemShut
  {NoStop}%
\bibitem [{\citenamefont {Hartle}(1967)}]{Hartle:1967he}%
  \BibitemOpen
  \bibfield  {author} {\bibinfo {author} {\bibfnamefont {J.~B.}\ \bibnamefont
  {Hartle}},\ }\href {https://doi.org/10.1086/149400} {\bibfield  {journal}
  {\bibinfo  {journal} {Astrophys. J.}\ }\textbf {\bibinfo {volume} {150}},\
  \bibinfo {pages} {1005} (\bibinfo {year} {1967})}\BibitemShut {NoStop}%
%%CITATION = ASJOA,150,1005;%%
\bibitem [{\citenamefont {{Bowers}}\ and\ \citenamefont
  {{Liang}}(1974)}]{1974ApJ...188..657B}%
  \BibitemOpen
  \bibfield  {author} {\bibinfo {author} {\bibfnamefont {R.~L.}\ \bibnamefont
  {{Bowers}}}\ and\ \bibinfo {author} {\bibfnamefont {E.~P.~T.}\ \bibnamefont
  {{Liang}}},\ }\href {https://doi.org/10.1086/152760} {\bibfield  {journal}
  {\bibinfo  {journal} {Astrophys. J.}\ }\textbf {\bibinfo {volume} {188}},\
  \bibinfo {pages} {657} (\bibinfo {year} {1974})}\BibitemShut {NoStop}%
\bibitem [{\citenamefont {Stephani}\ \emph {et~al.}(2003)\citenamefont
  {Stephani}, \citenamefont {Kramer}, \citenamefont {MacCallum}, \citenamefont
  {Hoenselaers},\ and\ \citenamefont {Herlt}}]{Stephani:2003tm}%
  \BibitemOpen
  \bibfield  {author} {\bibinfo {author} {\bibfnamefont {H.}~\bibnamefont
  {Stephani}}, \bibinfo {author} {\bibfnamefont {D.}~\bibnamefont {Kramer}},
  \bibinfo {author} {\bibfnamefont {M.~A.~H.}\ \bibnamefont {MacCallum}},
  \bibinfo {author} {\bibfnamefont {C.}~\bibnamefont {Hoenselaers}},\ and\
  \bibinfo {author} {\bibfnamefont {E.}~\bibnamefont {Herlt}},\ }\href
  {https://doi.org/10.1017/CBO9780511535185} {\emph {\bibinfo {title} {{Exact
  solutions of Einstein's field equations}}}},\ Cambridge Monographs on
  Mathematical Physics\ (\bibinfo  {publisher} {Cambridge Univ. Press},\
  \bibinfo {address} {Cambridge},\ \bibinfo {year} {2003})\BibitemShut
  {NoStop}%
%%CITATION = INSPIRE-619666;%%
\bibitem [{\citenamefont {{Lema{\^\i}tre}}\ and\ \citenamefont
  {{MacCallum}}(1997)}]{1997GReGr..29..641L}%
  \BibitemOpen
  \bibfield  {author} {\bibinfo {author} {\bibfnamefont {G.~A.}\ \bibnamefont
  {{Lema{\^\i}tre}}}\ and\ \bibinfo {author} {\bibfnamefont {M.~A.~H.}\
  \bibnamefont {{MacCallum}}},\ }\href
  {https://doi.org/10.1023/A:1018855621348} {\bibfield  {journal} {\bibinfo
  {journal} {General Relativity and Gravitation}\ }\textbf {\bibinfo {volume}
  {29}},\ \bibinfo {pages} {641} (\bibinfo {year} {1997})}\BibitemShut
  {NoStop}%
\bibitem [{\citenamefont {{DeBenedictis}}\ \emph {et~al.}(2006)\citenamefont
  {{DeBenedictis}}, \citenamefont {{Horvat}}, \citenamefont {{Iliji{\'c}}},
  \citenamefont {{Kloster}},\ and\ \citenamefont
  {{Viswanathan}}}]{eosgravastar}%
  \BibitemOpen
  \bibfield  {author} {\bibinfo {author} {\bibfnamefont {A.}~\bibnamefont
  {{DeBenedictis}}}, \bibinfo {author} {\bibfnamefont {D.}~\bibnamefont
  {{Horvat}}}, \bibinfo {author} {\bibfnamefont {S.}~\bibnamefont
  {{Iliji{\'c}}}}, \bibinfo {author} {\bibfnamefont {S.}~\bibnamefont
  {{Kloster}}},\ and\ \bibinfo {author} {\bibfnamefont {K.~S.}\ \bibnamefont
  {{Viswanathan}}},\ }\href {https://doi.org/10.1088/0264-9381/23/7/007}
  {\bibfield  {journal} {\bibinfo  {journal} {Classical and Quantum Gravity}\
  }\textbf {\bibinfo {volume} {23}},\ \bibinfo {pages} {2303} (\bibinfo {year}
  {2006})},\ \Eprint {https://arxiv.org/abs/gr-qc/0511097} {arXiv:gr-qc/0511097
  [gr-qc]} \BibitemShut {NoStop}%
\bibitem [{\citenamefont {Chirenti}\ and\ \citenamefont
  {Rezzolla}(2007)}]{chirenti2007tell}%
  \BibitemOpen
  \bibfield  {author} {\bibinfo {author} {\bibfnamefont {C.~B.}\ \bibnamefont
  {Chirenti}}\ and\ \bibinfo {author} {\bibfnamefont {L.}~\bibnamefont
  {Rezzolla}},\ }\href@noop {} {\bibfield  {journal} {\bibinfo  {journal}
  {Classical and Quantum Gravity}\ }\textbf {\bibinfo {volume} {24}},\ \bibinfo
  {pages} {4191} (\bibinfo {year} {2007})}\BibitemShut {NoStop}%
\bibitem [{\citenamefont {{Bisnovatyj-Kogan}}\ and\ \citenamefont
  {{Zel'dovich}}(1969)}]{1969Afz.....5..223B}%
  \BibitemOpen
  \bibfield  {author} {\bibinfo {author} {\bibfnamefont {G.~S.}\ \bibnamefont
  {{Bisnovatyj-Kogan}}}\ and\ \bibinfo {author} {\bibfnamefont {Y.~B.}\
  \bibnamefont {{Zel'dovich}}},\ }\href@noop {} {\bibfield  {journal} {\bibinfo
   {journal} {Astrofizika}\ }\textbf {\bibinfo {volume} {5}},\ \bibinfo {pages}
  {223} (\bibinfo {year} {1969})}\BibitemShut {NoStop}%
\bibitem [{\citenamefont {{Bisnovatyi-Kogan}}\ and\ \citenamefont
  {{Thorne}}(1970)}]{1970ApJ...160..875B}%
  \BibitemOpen
  \bibfield  {author} {\bibinfo {author} {\bibfnamefont {G.~S.}\ \bibnamefont
  {{Bisnovatyi-Kogan}}}\ and\ \bibinfo {author} {\bibfnamefont {K.~S.}\
  \bibnamefont {{Thorne}}},\ }\href {https://doi.org/10.1086/150478} {\bibfield
   {journal} {\bibinfo  {journal} {\apj}\ }\textbf {\bibinfo {volume} {160}},\
  \bibinfo {pages} {875} (\bibinfo {year} {1970})}\BibitemShut {NoStop}%
\bibitem [{\citenamefont {{Chandrasekhar}}(1972)}]{1972grec.conf..185C}%
  \BibitemOpen
  \bibfield  {author} {\bibinfo {author} {\bibfnamefont {S.}~\bibnamefont
  {{Chandrasekhar}}},\ }in\ \href@noop {} {\emph {\bibinfo {booktitle} {General
  Relativity}}},\ \bibinfo {editor} {edited by\ \bibinfo {editor}
  {\bibfnamefont {L.}~\bibnamefont {{O'Raifeartaigh}}}}\ (\bibinfo {year}
  {1972})\ pp.\ \bibinfo {pages} {185--199}\BibitemShut {NoStop}%
\bibitem [{\citenamefont {Chavanis}(2008)}]{Chavanis:2007kn}%
  \BibitemOpen
  \bibfield  {author} {\bibinfo {author} {\bibfnamefont {P.-H.}\ \bibnamefont
  {Chavanis}},\ }\href {https://doi.org/10.1051/0004-6361:20078287} {\bibfield
  {journal} {\bibinfo  {journal} {Astron. Astrophys.}\ }\textbf {\bibinfo
  {volume} {483}},\ \bibinfo {pages} {673} (\bibinfo {year} {2008})},\ \Eprint
  {https://arxiv.org/abs/0707.2292} {arXiv:0707.2292 [astro-ph]} \BibitemShut
  {NoStop}%
%%CITATION = ARXIV:0707.2292;%%
\bibitem [{\citenamefont {Ibohal}(2005)}]{Ibohal:2004kk}%
  \BibitemOpen
  \bibfield  {author} {\bibinfo {author} {\bibfnamefont {N.}~\bibnamefont
  {Ibohal}},\ }\href {https://doi.org/10.1007/s10714-005-0002-6} {\bibfield
  {journal} {\bibinfo  {journal} {Gen. Rel. Grav.}\ }\textbf {\bibinfo {volume}
  {37}},\ \bibinfo {pages} {19} (\bibinfo {year} {2005})},\ \Eprint
  {https://arxiv.org/abs/gr-qc/0403098} {arXiv:gr-qc/0403098} \BibitemShut
  {NoStop}%
\bibitem [{\citenamefont {{Dymnikova}}(2006)}]{2006PhLB..639..368D}%
  \BibitemOpen
  \bibfield  {author} {\bibinfo {author} {\bibfnamefont {I.}~\bibnamefont
  {{Dymnikova}}},\ }\href {https://doi.org/10.1016/j.physletb.2006.06.035}
  {\bibfield  {journal} {\bibinfo  {journal} {Physics Letters B}\ }\textbf
  {\bibinfo {volume} {639}},\ \bibinfo {pages} {368} (\bibinfo {year}
  {2006})},\ \Eprint {https://arxiv.org/abs/hep-th/0607174}
  {arXiv:hep-th/0607174 [hep-th]} \BibitemShut {NoStop}%
\bibitem [{\citenamefont {{Gonzalez de Urreta}}\ and\ \citenamefont
  {{Socolovsky}}(2015)}]{deUrreta:2015nla}%
  \BibitemOpen
  \bibfield  {author} {\bibinfo {author} {\bibfnamefont {E.~J.}\ \bibnamefont
  {{Gonzalez de Urreta}}}\ and\ \bibinfo {author} {\bibfnamefont
  {M.}~\bibnamefont {{Socolovsky}}},\ }\href@noop {} {\bibfield  {journal}
  {\bibinfo  {journal} {arXiv e-prints}\ ,\ \bibinfo {eid} {arXiv:1504.01728}}
  (\bibinfo {year} {2015})},\ \Eprint {https://arxiv.org/abs/1504.01728}
  {arXiv:1504.01728 [gr-qc]} \BibitemShut {NoStop}%
\bibitem [{\citenamefont {{Wolfram Research, Inc.}}()}]{Mathematica}%
  \BibitemOpen
  \bibfield  {author} {\bibinfo {author} {\bibnamefont {{Wolfram Research,
  Inc.}}},\ }\href {https://www.wolfram.com/mathematica} {\bibinfo {title}
  {Mathematica, {V}ersion 12.2.0}},\ \bibinfo {note} {{Champaign, IL,
  2020}}\BibitemShut {NoStop}%
\end{thebibliography}%

\end{document}